\documentclass[11pt,a4paper]{article}
\pdfoutput=1
\usepackage{jheppub}
\usepackage{graphicx}  
\usepackage{dcolumn}   
\usepackage{bm}        
\usepackage{amsmath,amssymb,amsxtra,bm,epsfig}   
\usepackage{float}
\usepackage{xcolor}
\usepackage{xspace}

\usepackage[font=scriptsize]{subcaption}

\title{ Flavored Leptogenesis and Neutrino Mass with $A_4$ symmetry}

\author[a]{Arghyajit Datta,}
\emailAdd{datta176121017@iitg.ac.in}
\affiliation[a]{Department of Physics, Indian Institute of Technology Guwahati, Assam-781039, India}

\author[b,c]{Biswajit Karmakar,}
\emailAdd{biswajit.karmakar@us.edu.pl}
\affiliation[b]{Department of Physics, Indian Institute of Technology Hyderabad, 502285 Telangana, India} 
\affiliation[c]{Institute of Physics, University of Silesia, 40-007 Katowice, Poland}

\author[a]{Arunansu Sil}
\emailAdd{asil@iitg.ac.in}

\abstract{We propose a minimal $A_4$ flavor symmetric model, assisted by $Z_2 \times Z_3$ symmetry, which can naturally takes care of the appropriate lepton mixing and neutrino masses via Type-I seesaw. It turns out that the framework, originated due to a specific flavor structure, favors the normal hierarchy of light neutrinos and simultaneously narrows 
	down the range of Dirac CP violating phase. It predicts an interesting correlation between the atmospheric mixing angle and the Dirac CP phase too. While the flavor structure indicates an exact degeneracy of the right-handed neutrino masses, renormalization group running of the same from a high scale is shown to make it quasi-degenerate and a successful flavor leptogenesis takes place within the allowed parameter space obtained from neutrino phenomenology.  
}
\arxivnumber{2106.06773}
\begin{document}
	\maketitle
	\flushbottom
	\section{Introduction}\label{sec:intro}  
	
	Over the last couple of decades, we have witnessed remarkable success in neutrino experiments~\cite{Fukuda:2001nk, Fukuda:2002pe, Ashie:2005ik, Ahmad:2002jz, Ahmad:2002ka, Abe:2008aa, Abe:2011sj, Abe:2011fz, Abe:2013hdq} indicating that neutrinos are indeed massive. Furthermore, mixing parameters have been measured with great precision. In fact, two of the three mixing angles namely solar ($\theta_{12} $) and atmospheric ($\theta_{23}$) ones are found to be large while the other one, reactor ($\theta_{13}$) mixing angle, is relatively small. Such a finding clearly shows the distinctive feature associated to the lepton sector in contrast to the quark one where all the three mixing angles are measured to be small. 
	To have a deeper understanding of it, one needs to investigate the origin of the neutrino mass by looking at the neutrino mass matrix as well as the charged lepton sector from a symmetry perspective. 
	
	To address the tiny neutrino mass issue, various seesaw mechanisms~\cite{Minkowski:1977sc, GellMann:1980vs, Mohapatra:1979ia, Mohapatra:1986aw, Magg:1980ut, Lazarides:1980nt, Schechter:1980gr, Mohapatra:1986bd} have been proposed 
	by extending the Standard Model (SM) with heavy fermions/scalars. Among these, the type-I seesaw mechanism provides perhaps the simplest explanation of tiny neutrino mass where the SM is extended by three singlet right-handed neutrinos (RHN)~\cite{Minkowski:1977sc, GellMann:1980vs, Mohapatra:1979ia}. Involvement of flavor symmetries within this simple setup is off course an interesting possibility in order to explain the typical mixing pattern in the lepton sector. Non-abelian discrete groups (like $S_3, A_4, S_4, A_5, \Delta(27)$ etc.) in this regard have been extensively used (see reviews~\cite{Altarelli:2010gt, Xing:2019vks, King:2015aea, Grimus:2011fk, King:2017guk, King:2011zj, King:2013eh, Ishimori:2010au} and references therein).  
	
	Among the various discrete groups, $A_4$ turns out to be the most economical one\footnote{$A_4$ group was initially proposed as an underlying family symmetry for quark sector by ~\cite{Wyler:1979fe,Branco:1980ax}.}. It is a group of even permutations of four objects having three inequivalent one-dimensional representations (1, $1'$ and $1''$) as well as one three-dimensional representation (3). Interestingly, the three generations (or flavors) of right-handed charged lepton singlets 
	can naturally fit into these three inequivalent one-dimensional representations of $A_4$ while the three SM lepton doublets can be accommodated into the triplet representation of $A_4$~\cite{Altarelli:2005yx, Altarelli:2005yp, Ma:2001dn}. 
	Works along this direction ~\cite{Altarelli:2005yp, Altarelli:2005yx, Ma:2004zv, Altarelli:2009kr} showed that type-I seesaw model with $A_4$ flavor symmetry in general leads to a typical tri-bimaximal (TBM) lepton mixing ($\sin^2\theta_{12} = 1/3, ~\sin^2\theta_{23} = 1/2$ and $\theta_{13}=0$) pattern~\cite{Harrison:2002er, Harrison:2002kp} in presence of SM singlet (though charged under $A_4$) flavon fields. Though such TBM pattern received a great deal of attention due to its close proximity with experimental observation prior to 2012, it fails to accommodate the recent observation of small, but non-zero 
	$\theta_{13}$~\cite{An:2012eh, Abe:2011fz, Ahn:2012nd}. Subsequently, modifications over models based on $A_4$ (and other discrete groups) are suggested to accommodate non-zero $\theta_{13}$ either by considering additional flavon fields 
	or including corrections to vaccuum alignments of the flavons \cite{Branco:2012vs, Li:2016nap} or considering contributions 
	to additional mixing from the charged lepton sector~\cite{Araki:2010ur}. 
	
	In this work, we particularly focus on a framework where a non-trivial contribution to lepton mixing is originated from charged lepton sector. We do not consider any additional flavon field apart from those ones incorporated in the original Altarelli-Feruglio (AF) model \cite{Altarelli:2005yx}. While the RHN mass matrix turns out to be diagonal as a result of the flavor symmetry imposed, the structure of the charged lepton mass matrix becomes such that it can be diagonalized by a complex `magic' matrix \cite{Altarelli:2005yp}. Interestingly, an antisymmetric contribution to the Dirac neutrino mass matrix, originated from the product of two $A_4$ triplets, plays a crucial role in generating non-zero $\theta_{13}$~\cite{Memenga:2013vc, Borah:2017dmk, Borah:2018gjk, Borah:2018nvu} in our model which was overlooked in an earlier attempt~\cite{Branco:2009by}. In doing the analysis, we 
	find the atmospheric mixing angle $\theta_{23} \leq 45^{\circ} ~ i.e.$ to lie in lower octant (LO). We also note that only normal hierarchy (NH) of light neutrino masses are allowed in this model. This turns out to be another salient feature of our construction. 
	These predictions can be tested in ongoing and future neutrino experiments as ambiguities are still present in determining octant for $\theta_{23}$ as well as hierarchies of light neutrino masses. 
	
	Additionally, we also discuss the aspects of leptogenesis ~\cite{Fukugita:1986hr, Luty:1992un, Plumacher:1996kc, Covi:1996wh,Datta:2021elq} from the CP-violating decays of RHNs in this $A_4$ based type-I seesaw scenario in 
	line with observations \cite{Borah:2017qdu, Karmakar:2014dva, Karmakar:2015jza, Bhattacharya:2016lts, Bhattacharya:2016rqj, Hagedorn:2009jy, Jenkins:2008rb,Das:2019ntw}. In doing so, since the involvement of the neutrino Yukawa matrix in the charged-lepton mass diagonal basis is necessary, the specific flavor symmetric construction of it is expected to play an important role. In fact, due to this symmetry, exactly degenerate heavy RHNs result at tree level, thereby indicating the breaking of the perturbative field theory involved in CP asymmetry generation  \cite{Pilaftsis:1997jf}. Following \cite{Branco:2009by}, we are able to show that running of the parameters involved in the neutrino sector from the flavor symmetry breaking scale to the RHN mass scale actually eliminates such exact degeneracies  
	and as a result, leptogenesis can indeed be possible. The present study of matter-antimatter asymmetry generation via leptogenesis taking into account the effect of running however differs from that of \cite{Branco:2009by} by two aspects. Firstly, we use less number of flavon fields and secondly, we present a detailed analysis of flavored leptogenesis by solving the relevant Boltzmann equations. 
	
	The rest of the paper is organized as follows. In Section \ref{sec:structure} we present detail structure of the model including the analysis of the mixing matrices involved. Section \ref{sec:pheno} deals with phenomenology of neutrino mixing. Constrains and predictions on neutrino parameters (including neutrinoless double beta decay) involved are presented here. In Section \ref{sec:lepto} we perform a detailed study on leptogenesis solving flavored Boltzmann equations. Finally in Section \ref{sec:conc} we summarize the results and make final conclusion.

	\section{Structure of The Model}\label{sec:structure}
	
	 To realize the canonical type-I seesaw mechanism, we first consider an extension of the SM by including three singlet RHN fields ($N_R$).  Additionally, three flavon fields namely $\Phi$, $\Psi$, $\varphi$ and a discrete symmetry $A_4 \times Z_2\times Z_3$ are also incorporated to probe the typical flavor structure involved in the lepton sector. Note that same fields content was also present in the original AF \cite{Altarelli:2010gt} construction. Here $N_R$ and the flavon fields $\Phi$, $\Psi$ transform as triplet, whereas $\varphi$ transforms as a singlet under $A_4$. A judicious choice of additional $Z_2 \times Z_3$ symmetry assists the leptonic mass matrices to take specific forms and hence forbid several unwanted contributions. In Table \ref{tab:fields}, we present transformation properties of all the relevant SM fields, $N_R$ and flavons involved in the analysis.
	\begin{table}[h]
		\begin{center}
			\begin{tabular}{|c||c|c|c|c|c||c|c|c|c|c|}
				\hline
				{\tt Fields} & $\ell$ & $e_R$ & $\mu_R$ & $\tau_R$ & $N_R$ & $ H $& $\varphi$ & $\Phi$ &  $\Psi $ \\
				\hline
				SM & $(2,1/2)$ & $(1,1)$ & $(1,1)$ & $(1,1)$ & $(1,0)$ & $(2,-1/2)$ & $(1,0)$ & $(1,0)$ & $(1,0)$ \\
				\hline
				$A_4$ & $3$ & $1$ & $1^\prime$ & $ 1^{\prime\prime}$ & $3$ & $1$ & $1 $ &$3$ & $3$ \\
				\hline
				$Z_2$ & $1$ & $1$ & $1$ & $1$ & $-1$ & $1$ & $-1$ & $1$ & $-1$\\
				\hline
				$Z_3$ & $\omega$ & $1$ & $1$ & $1$ & $1$ & $1$ & $\omega$ & $\omega  $ & $\omega$\\
				\hline
			\end{tabular}
			\caption{\label{reps} Representations of the fields under $ SU(2)_L \times U(1)_Y\times A_4 \times Z_2\times Z_3$ symmetry}\label{tab:fields}
		\end{center}
	\end{table}
	
	The relevant effective Lagrangian involving charged leptons and neutrinos can be written as 
	\begin{equation}\label{eq:lag1}
	\begin{split}
	\mathcal{L} \supset &\frac{y_1^\ell}{\Lambda}\left(\bar{\ell}\,\Phi\right)_{\mathbf{1}}H\,e_R+
	\frac{y_2^\ell}{\Lambda}\left(\bar{\ell}\,\Phi\right)_{\mathbf{1^{\prime\prime}}}H\,\mu_R + \frac{y_3^\ell}{\Lambda}\left(\bar{\ell}\,\Phi\right)_{\mathbf{1^\prime}}H\, \tau_R \\& + \frac{y_1^\nu}{\Lambda}\left[(\bar{\ell}\,N_R)_{{\bf s}}\,\Psi \right]_{\mathbf{1}}\tilde{H} +\frac{y_2^\nu}{\Lambda}\left[(\bar{\ell}\,N_R)_{{\bf a}}\,\Psi\right]_{\mathbf{1}}\tilde{H}+\frac{y_3^\nu}{\Lambda}\left(\bar{\ell}\,N_R\right)_{\mathbf{1}}\varphi\,\tilde{H} +\frac12 M \left (\overline{N^c_R} N_R\right) + \text{h.c.},
	\end{split}
	\end{equation}
	where $y_{i=1,2,3}^{\ell,\nu}$ are the respective coupling constants, $M$ is the mass parameter of RHNs 
	and $\Lambda$ is the cut-off scale of the theory. In the first line of Eq. (\ref{eq:lag1}), terms in the first parentheses represent products of two $A_4$ triplets forming a one-dimensional representation which further contract 
		with 1, $1'$ and $1''$ of $A_4$, corresponding to $e_R$, $\mu_R$ and $\tau_R$ respectively, to make a true singlet under $A_4$. On the other hand, in the second line of Eq. (\ref{eq:lag1}), the subscripts ${\bf s}$, ${\bf a}$ correspond to symmetric and anti-symmetric parts of triplet products in the $S$ diagonal basis of $A_4$. The essential multiplication rules of the $A_4$ group elements are elaborated in appendix~\ref{apa}.  
	
	From Table \ref{tab:fields} it is evident that the tree level contribution to charged lepton Yukawa interaction, $\bar{\ell}\,H \alpha_{R}$ (with $\alpha=e,\mu,\tau$), gets forbidden. Instead, such interactions are effectively generated once the flavon 
	$\Phi$ gets a vacuum expectation value (vev) via the dimension-5 operators (present in first line of Eq. (\ref{eq:lag1})). 
	Similarly in the neutrino sector, the renormalizable Dirac Yukawa coupling is forbidden as the lepton doublet $\ell$ is charged under $Z_3$ whereas both $N_R$ and $H$ transform trivially under it. However such effective Yukawa coupling is generated from dimension-5 operators involving flavons 
	$\Psi$ and $\varphi$, after they obtain vevs. Presence of $Z_2$ symmetry is important in identifying $\Phi$ from $\Psi$ (both being $A_4$ triplet) so that they contribute to the charged lepton and Dirac neutrino Yukawa couplings differently.

	The flavon fields break the flavor symmetry $A_4 \times Z_3 \times Z_2$ when they acquire vevs along{\footnote{Such vev alignments of the flavons is widely used and  can be realised in a natural way by minimising the scalar potential following the  approach of \cite{Altarelli:2010gt, Karmakar:2016cvb, Memenga:2013vc, He:2006dk, Lin:2008aj, Branco:2009by, Rodejohann:2015hka}.}
		\begin{equation} \label{vevalign}
		\begin{array}{ccc}
		\left<\varphi\right>=v_\varphi \,,\quad&\left<\Phi\right>=v_\Phi \left (1,1,1\right)\,,
		\quad&\left<\Psi \right>=v_\Psi\left(0,1,0\right),\,
		\end{array}
		\end{equation}
		as a result of which the part of the Lagrangian contributing  to the charged lepton sector can be written as
		\begin{equation}
		\mathcal{L}_l = \frac{y_1^\ell v_\Phi}{\Lambda}(\bar{\ell}_e+\bar{\ell}_\mu +\bar{\ell}_\tau) H\, e_R  + \frac{y_2^\ell v_\Phi}{\Lambda}(\bar{\ell}_e+\omega\bar{\ell}_\mu +\omega^2\bar{\ell}_\tau)  H \, \mu_R + \frac{y_3^\ell v_\Phi}{\Lambda}(\bar{\ell}_e+\omega^2\bar{\ell}_\mu +\omega\bar{\ell}_\tau) H\, \tau_R. 
		\end{equation}
		Using the above Lagrangian one obtains the charged lepton mass matrix after the electroweak symmetry breaking as
			\begin{align}\label{yl}
			Y^\ell=v 
			\begin{pmatrix}
			f_1^\ell&f_2^\ell&f_3^\ell \\
			f_1^\ell&\omega\,f_2^\ell&\omega^2\,f_3^\ell\\
			f_1^\ell&\omega^2\,f_2^\ell&\omega\,f_3^\ell
			\end{pmatrix};
			~~f_i^\ell=\frac{v_\Phi}{\Lambda}\,y_i^\ell ~{\rm{with}}~i=1,2,3,
			\end{align}
			where $v$= 174 GeV stands for the vev of the SM Higgs. 
		
		In a similar way, the Lagrangian for neutrino sector after breaking of the flavor symmetries can be written as 
		\begin{equation}
		\begin{split}
		\mathcal{L}_\nu=&\frac{y_3^\nu}{\Lambda}\,v_\varphi\left(\bar{\ell_e}\,N_{1R}+
		\bar{\ell_\mu}\,N_{2R} +  \bar{\ell_\tau}\,N_{3R}\right)\,\tilde{H}+ 
		(y_1^\nu-y_2^\nu)\,\frac{v_\Psi}{\Lambda}\bar{\ell_e}\,N_{3R}\, \tilde{H}\\
		&+
		(y_1^\nu+y_2^\nu)\,\frac{v_\Psi}{\Lambda}\bar{\ell_\tau}\,N_{1R}\, \tilde{H}
		+ M\left(\overline{N^c}_{1R} N_{1R}+ \overline{N^c}_{2R} N_{2R}+
		\overline{N^c}_{3R}N_{3R}\right)+\text{h.c.}.
		\end{split}
		\end{equation}
		This yields the corresponding Dirac and Majorana mass matrices as
		\begin{align}\label{Ynudef}
		Y^{\nu}=
		\begin{pmatrix}
		f_3^\nu & 0 &f_1^\nu-f_2^\nu\\
		0 & f_3^\nu & 0\\
		f_1^\nu+f_2^\nu & 0 & f_3^\nu
		\end{pmatrix},
		\end{align}
		
		\begin{align}\label{MR}
		M_R =\begin{pmatrix}
		M& 0 &0\\
		0&M&0\\
		0&0&M
		\end{pmatrix},
		\end{align}
		with $f_i^\nu=\frac{ \,v_\Psi}{\Lambda} y_i^\nu$, $i=1,2,3$.
		
		Let us now discuss the diagonalization of the charged lepton and neutrino mass matrices so as to obtain the lepton mixing matrix. First we note that the charged lepton mass matrix given in Eq. (\ref{yl}) can be diagonalized by a bi-unitary transformation 
		\begin{align}\label{yldiag}
		Y^\ell=  Vd^\ell \mathbb{I}_{3\times 3} ~~{\rm with}~~d^\ell=\sqrt{3} v\,\text{diag}\,(f_1^\ell,f_2^\ell,f_3^\ell),
		\end{align}	
		where $\mathbb{I}_{3\times 3}$ is a $3\times 3$ identity matrix and
		\begin{align}
		V=\frac{1}{\sqrt{3}}
		\begin{pmatrix}
		1&1&1\\
		1&\omega&\omega^2\\
		1&\omega^2&\omega
		\label{Vmatrix}
		\end{pmatrix}, 
		\end{align}
		where $\omega$ ($=e^{2i\pi/3}$) is the cube root of unity. From Eq. (\ref{MR}), it is evident  that the right-handed Majorana neutrino mass matrix $M_R$ is diagonal having degenerate mass eigenvalues ($M$) to start with. 
		
		On the other hand, $f_1^\nu$ and $f_2^\nu$ appearing in Eq. (\ref{Ynudef}) are the symmetric and antisymmetric contributions to the Dirac neutrino Yukawa respectively, originated as products of two $A_4$ triplets $\ell$ and $N_R$ which further contract with $\Phi$ (see the product rules Eq. (\ref{eq:3s}) and (\ref{eq:3a}). 
			This antisymmetric part plays an instrumental role\footnote{Earlier the role of such antisymmetric contributions was analyzed in the context of Dirac neutrinos \cite{Memenga:2013vc, Borah:2017dmk, Borah:2018gjk, Borah:2018nvu}.} in realizing correct neutrino oscillation data.

		Here it is worth mentioning that in the vanishing limit of $f_2^\nu \rightarrow 0$, (keeping the structure of the charged lepton and Majorana mass matrix intact) one can reproduce the TBM mixing as discussed in \cite{Branco:2009by}.
		
		The effective light neutrino mass\footnote{With the symmetries mentioned in Table \ref{reps} in principle,  there will be a contribution to the effective light neutrino mass via a dim-6 operator given by $\frac{y_{eff}}{\Lambda^2} (\ell H \ell H \Phi)$. However, in the limit $v_{\Phi}>M$,  this additional contribution can be neglected compared to the  dominant  type-I contribution considered here.} matrix can be obtained within the type-I seesaw framework as 
		\begin{align}\label{mnuseesaw}
		m_\nu= -m_D M_R^{-1}m_D^T, 
		\end{align}
		where the structure of $M_R$ is  given in Eq. (\ref{MR}).  Now, from Eq. (\ref{Ynudef}) and (\ref{yldiag}),  in the basis where the charged leptons are diagonal, the Dirac neutrino mass matrix in that basis can be written as,	
		\begin{align}\label{diagMD}
		m_D= v V^\dagger Y^{\nu}= v \mathcal{Y}^{\nu}. 
		\end{align}
		Therefore, substituting Eq. (\ref{diagMD}) in the type-I seesaw formula given by  Eq. (\ref{mnuseesaw}) one obtains the light neutrino mass matrix as  
		\begin{align}\label{mnuynuynut}
		m_\nu=&- v^2 V^\dagger Y^{\nu} M_R^{-1}{Y^{\nu}}^T V^*,\\
		&=-\frac{1}{M}V^\dagger ( v^2 Y^{\nu} {Y^{\nu}}^T )V^*\label{mnuynuynut2}. 
		\end{align}
		
		Clearly, to get the mass eigenvalues of light neutrinos we need to diagonalize $Y^{\nu} {Y^{\nu}}^T$ where
		
		\begin{align}\label{ynuynut}
		{Y^{\nu}} {Y^{\nu}}^T=  \left(
		\begin{array}{ccc}
		(f_1^\nu-f_2^\nu)^2+f_3^{\nu^2} & 0 & 2f_1^\nu f_3^\nu \\
		0 & f_3^{\nu^2} & 0 \\
		2f_1^\nu f_3^\nu & 0 & (f_1^\nu+f_2^\nu)^2+f_3^{\nu^2}\\
		\end{array}
		\right).
		\end{align}
		{Though $Y^{\nu}$ is in general a complex matrix,  $Y^{\nu} {Y^{\nu}}^T$ being a complex symmetric matrix can be diagonalized by an orthogonal transformation (in the $(1,3)$ plane) through the relation 
		\begin{align}\label{ynud}
		U_{13}^T (Y^{\nu} {Y^{\nu}}^T) U_{13} = d_D^2= \text{diag}(\lambda_1,\lambda_2,\lambda_3),
		\end{align}
		where the rotation matrix $U_{13}$ (parametrised by angle $\theta$ and phase $\psi$) is given by 
		\begin{align}\label{u13}
		U_{13}=\left(
		\begin{array}{ccc}
		\cos \theta & 0 & e^{-i \psi } \sin \theta  \\
		0 & 1 & 0 \\
		-e^{i \psi } \sin \theta  & 0 & \cos \theta  \\
		\end{array}
		\right).
		\end{align}
		The complex eigenvalues are given by 
		\begin{align}
		\lambda_1 &= f_1^{\nu^2}+f_2^{\nu^2}+f_3^{\nu^2}-2\sqrt{f_1^{\nu^2}(f_2^{\nu^2}+f_3^{\nu^2})},\\
		\lambda_2 &= f_3^{\nu^2},\\
		\lambda_3 &=f_1^{\nu^2}+f_2^{\nu^2}+f_3^{\nu^2}+2\sqrt{f_1^{\nu^2}(f_2^{\nu^2}+f_3^{\nu^2})}.
		\end{align}
		Now substituting Eq. (\ref{ynud}) in Eq. (\ref{mnuynuynut2}), we get 
		\begin{align}
		m_\nu &= -V^\dagger U_{13} \left( \frac{v^2 d_D^2}{M} \right)U_{13}^T V^*,\\
		&=  -V^\dagger U_{13} \left( d_\nu \right)U_{13}^T V^*\label{mnud},
		\end{align}
		where $d_\nu= v^2 d_D^2/M$ is a diagonal matrix  having diagonal elements $v^2 \lambda_i/M$ ($i=1,2,3$), representative of three complex light neutrino mass eigenvalues.  	
		
		In order to extract the real and positive light neutrino mass eigenvalues, we choose the following representations of the 
		parameters $f_{1,2,3}^\nu (=|f_{1,2,3}^{\nu}|e^{i\phi_{1,2,3}}$ and $\phi_{1,2,3}$ are the three phases associated) as
		\begin{align}
		\frac{f_1^\nu}{f_3^\nu}=\frac{\mid f_1^\nu \mid}{\mid f_3^\nu \mid} e^{ i (\phi_1-\phi_3)}=\chi_1 e^{i\gamma_{1}} \label{chi1},\\
		\frac{f_2^\nu}{f_3^\nu}=\frac{\mid f_2^\nu \mid}{\mid f_3^\nu \mid} e^{ i (\phi_2-\phi_3)}=\chi_2 e^{i\gamma_{2}} \label{chi2},
		\end{align}
		where $\chi_1=|f_1^\nu/f_3^\nu|$, $\chi_2=|f_2^\nu/f_3^\nu|$ and $(\phi_1- \phi_3)=\gamma_1$, 
			$(\phi_2- \phi_3)=\gamma_2$ are the redefined parameters used for the rest of our analysis. 
		
		Now we are in a position to define the rotation angle $\theta$ and phase $\psi$ of $U_{13}$ matrix (see Eq. (\ref{u13})) as:
		\begin{align}
		\tan2\theta&= \frac{2 \chi_1}{2\chi_1 \chi_2 \cos \gamma_{2} \cos \psi-\left[\chi_1^2 \sin \gamma_{1}+\chi_2^2 \sin (2 \gamma_{2}-\gamma_{1})-\sin \gamma_{1}\right]\sin \psi}\label{theta},\\
		\tan \psi&= \frac{-2 \chi_1 \chi_2 \sin \gamma_{2}}{\cos \gamma_{1}+\chi_2^2 \cos(2\gamma_{2}-\gamma_{1})+\chi_1^2 \cos \gamma_{1}}\label{psi}.
		\end{align}
		Similarly, the real and positive light neutrino masses can also be expressed in terms of $\chi_{1,2}$ and $\gamma_{1,2}$ after we extract the phases from the complex eigenvalues. To proceed, note that Eq. (\ref{mnud}) can be rewritten as 
		\begin{align}\label{mnuU}
		m_\nu \equiv U {\rm{diag}} (m_1, m_2, m_3) U^T,
		\end{align}
		with 
		\begin{align}\label{U}
		U= V^\dagger U_{13} e^{i \frac{\pi}{2}}U_p.
		\end{align}
		Here $U_p$ stands for a diagonal phase matrix given by $U_p={\rm diag} (1,e^{i\beta_{21}/2},e^{i\beta_{31}/2})$, and 
		the real positive light neutrino masses are given by:
		\begin{align}
		m_1 &= \frac{v^2}{M} \mid f_3^{\nu^2} \mid \sqrt{o_r^2+o_i^2},\label{m1}\\
		m_2 &= \frac{v^2}{M} \mid f_3^{\nu^2} \mid,\label{m2}\\
		m_3  &= \frac{v^2}{M} \mid f_3^{\nu^2} \mid \sqrt{n_r^2+n_i^2}\label{m3},
		\end{align}
		where $o_r, o_i, n_r$ and $n_i$ can be written in terms of the associated parameters ($\chi_1, \chi_2, \gamma_1$ and $\gamma_2$) in our model   as
		\begin{align}
		o_r &= \chi_1^2 \cos 2\gamma_{1}+\chi_2^2\cos 2\gamma_{2}+1 - 2 A \chi_1 \cos\gamma_{1}+ 2 B\chi_1 \sin \gamma_{1},\label{or}\\
		o_i &=\chi_1^2\sin 2\gamma_{1}+\chi_2^2 \sin 2\gamma_{2}-2 A \chi \sin \gamma_{1} -2 B \chi_1 \cos \gamma_{1},\\
		n_r &=\chi_1^2 \cos 2\gamma_{1}+\chi_2^2\cos 2\gamma_{2}+1 + 2 A \chi_1 \cos\gamma_{1}- 2 B\chi_1 \sin \gamma_{1},\\
		n_i &=\chi_1^2\sin 2\gamma_{1}+\chi_2^2 \sin 2\gamma_{2}+2 A \chi \sin \gamma_{1} +2 B \chi_1 \cos \gamma_{1}.\label{ni}, \\
		A &= \frac{\sqrt{1+\chi^2_2\cos{2\gamma_2}+\sqrt{1+\chi^4_2 +2 \chi^2_2 \cos 2\gamma_2}}}{\sqrt{2}},
		B =\frac{\chi_2^2 \sin 2\gamma_2}{2A}. 
		\end{align}
		The phases $\beta_{21(31)}$ involved in $U_{p}$ are given by,		
		\begin{align}	
		\beta_{21}&= -\tan ^{-1}\frac{o_i}{o_r},
		\beta_{31} =\tan ^{-1}\frac{n_i}{n_r}-\tan ^{-1}\frac{o_i}{o_r}. \label{B}
		\end{align}
		
		Therefore using Eqs. (\ref{Vmatrix}), (\ref{u13}) and (\ref{U}), the final form of the mixing matrix $U$ which diagonalises the effective light neutrino mass matrix (in the charged lepton diagonal basis) can now be written as 
		\begin{align}\label{ueff}
		U=
		\left(
		\begin{array}{ccc}
		\frac{\cos \theta - e^{i \psi } \sin \theta}{\sqrt{3}}  & \frac{1}{\sqrt{3}} & \frac{\cos \theta + e^{-i \psi } \sin \theta}{\sqrt{3}}   \\
		\frac{\cos \theta - \omega e^{i \psi } \sin \theta}{\sqrt{3}}  & \frac{\omega^2}{\sqrt{3}}  &\frac{\omega \cos \theta + e^{-i \psi } \sin \theta}{\sqrt{3}} \\
		\frac{\cos \theta - \omega^2 e^{i \psi } \sin \theta}{\sqrt{3}} &  \frac{\omega}{\sqrt{3}} & \frac{\omega^2 \cos \theta + e^{-i \psi } \sin \theta}{\sqrt{3}} \\
		\end{array}
		\right) e^{i\pi/2}U_p.
		\end{align}
		$U$ is therefore the lepton mixing matrix, called the Pontecorvo-Maki-Nakagawa-Sakata ($U_{PMNS}$) matrix, the standard form of which is given by \cite{Tanabashi:2018oca}, 
		\begin{align}\label{upmns}
		U_{PMNS}= 
		\begin{pmatrix}
		c_{12} c_{13} & s_{12} c _{13} & e^{-i \delta} s_{13}\\
		-s_{12}s_{23}- e^{i \delta} c_{12}s_{13}s_{23} & c_{12}c_{23}-e_{i \delta}s_{12}s_{13}s_{23} & c_{13}s_{23}\\
		s_{12}s_{23}- e^{i \delta} c_{12}s_{13}s_{23} & -c_{12}c_{23}-e_{i \delta}s_{12}s_{13}s_{23} & c_{13} c_{23}
		\end{pmatrix}
		U_m,
		\end{align}
		where $c_{ij}=\cos\theta_{ij}$ and  $s_{ij}=\sin\theta_{ij}$, and $\delta$ is the CP violating Dirac phase. Also, $U_m={\rm diag}(1, e^{i\alpha_{21}/2}, e^{i\alpha_{31}/2})$ is a  phase matrix which contains two Majorana phases  $\alpha_{21}$ and $\alpha_{31}$.
		Comparing above two matrices given in Eq. (\ref{ueff}) and (\ref{upmns}) we get the correlation between the neutrino mixing angles (and Dirac CP phase) appearing in $U_{PMNS}$ and the model parameters as~\cite{Memenga:2013vc}
		\begin{align}
		\mid s_{13} \mid ^2=& \frac{1+\sin2 \theta \cos \psi}{3} , \quad\tan\delta= \frac{\sin\theta \sin\psi}{\cos\theta +\sin\theta\cos\psi},\label{col1}\\
		s_{12}^2=& \frac{1}{3(1-\mid s_{13}\mid ^2)}, \quad\tan2\theta_{23} \cos\delta=\frac{1-2\mid s_{13}\mid ^2 }{\mid s_{13}\mid \sqrt{2-3 \mid s_{13}\mid^2}}\label{col2}. 
		\end{align}
		Additionally, the two Majorana phases $\alpha_{21}$ and $\alpha_{31}$ are  identified as 
			$\alpha_{21} = \beta_{21}$, and $\alpha_{31} = \beta_{31}$ (ignoring the irrelevant common phase).  	 
		These correlations given in Eq. (\ref{col1})-(\ref{col2}) are the keys to the subsequent analysis of neutrino phenomenology.

		\section{Neutrino phenomenology}\label{sec:pheno}
		\subsection{Constraining the parameter space}

		\begin{table}
			\centering
			\begin{tabular}{|c|c|c|}
				\hline
				parameters & best fit value & $3\sigma$ range\\ 
				\hline
				$\sin^2\theta_{12}$ & $0.304$ & $0.269 \to 0.343$
				\\
				\hline
				$\sin^2\theta_{23}$	& $0.573$ & $0.415 \to 0.616$
				\\
				\hline
				$\sin^2\theta_{13}$	& $0.02219$ & $0.02032 \to 0.02410$
				\\
				\hline
				$\delta_{CP}/^\circ$ & $197$ & $120 \to 369$
				\\
				\hline
				$\frac{\Delta m_{21}^2}{10^{-5}~eV^2}$	& $7.42$ & $6.82 \to 8.04$
				\\
				\hline
				$\frac{\Delta m_{31}^2}{10^{-3}~eV^2}$ & $+2.517$ & $+2.435 \to +2.598$
				
				\\
				\hline
			\end{tabular}
		
			\caption{neutrino oscillation data	 obtained from NuFIT\cite{Esteban:2020cvm} for NH scenario of light neutrino mass.}\label{nexp}
		\end{table}

		As seen from Eqs. (\ref{col1}) and (\ref{col2}) in conjugation with Eqs. (\ref{theta}) and (\ref{psi}), all the mixing angles 
		($\theta_{13}, \theta_{12}, \theta_{23}$) and the Dirac CP phase  $(\delta)$ involved in the lepton mixing matrix $U_{{PMNS}}$ are finally determined by the model parameters $\chi_1, \chi_2, \gamma_1$ and $\gamma_2$. Hence, 
			using the 3$\sigma$ allowed ranges of the three mixing angles ($\theta_{13}, \theta_{12}, \theta_{23}$) from neutrino oscillation data\footnote{The Majorana phases are insensitive to neutrino oscillation experiments. However, they may play an important role in neutrinoless double beta decay \cite{Langacker:1986jv}.} presented in Table \ref{nexp}, we can restrict parameter space for $\chi_{1,2}$ and $\gamma_{1,2}$. This parameter space of the current set-up can be further constrained using the 3$\sigma$ allowed ranges of the mass-squared differences (see Table \ref{nexp}). 
		For that purpose, we introduce a dimensionless quantity $r$, defined as the ratio of solar to atmospheric mass squared difference for normal hierarchy, $i.e.$, $r=\frac{\Delta m_{21}^2}{\Delta m_{31}^2}$ with $\Delta m_{21}^2=m_2^2-m_1^2$ and $\Delta m_{31}^2=m_3^2-m_1^2$. Using the three light neutrino mass eigenvalues  given in Eq. (\ref{m1})-(\ref{m3}), we are able to 
		rewrite it as
			\begin{align}\label{r}
			r=\frac{\Delta m_{21}^2}{\Delta m_{31}^2}= \frac{1-o_r^2-o_i^2}{ n_r^2 +n_i^2 -o_r^2-o_i^2 }.
			\end{align}
		Substituting $o_{r,i}, n_{r,i}$ from Eq. (\ref{or})-(\ref{ni}) into Eq. (\ref{r}), we note that $r$ now becomes function of $\chi_1, \chi_2, \gamma_1$ and $\gamma_2$. Apart from the satisfaction of $r$ value obtained from the ratio of the best fit values of mass-squared differences, we must satisfy both the individual mass-squared differences, $\Delta m_{21}^2$ and $\Delta m_{31}^2$, independently within their 3$\sigma$ allowed ranges using Eqs. \ref{m1}-\ref{m3}. There also exists a cosmological upper bound on sum of the light neutrinos masses as  $\sum_{i} m_i \leq 0.11$ eV~\cite{Vagnozzi:2017ovm,Aghanim:2018eyx} which will also constrain the parameter space. Note that in order to evaluate $\sum_{i} m_i$, we need to get an estimate of the pre-factor $|f_3^{\nu^2}|v^2/M$ (see Eqs. (\ref{m1})-(\ref{m3})) which can be obtained by using the relation 
		\begin{align}\label{prefac}
		|f_3^{\nu^2}|v^2/M=\sqrt{\Delta m_{21}^2/(1-o_r^2-o_i^2)},  
		\end{align}
		with the known value of $\Delta m_{21}^2$ from current global analysis~\cite{Esteban:2020cvm}. 
		
		\begin{figure}[h]
			$$
			\includegraphics[width=0.5\linewidth]{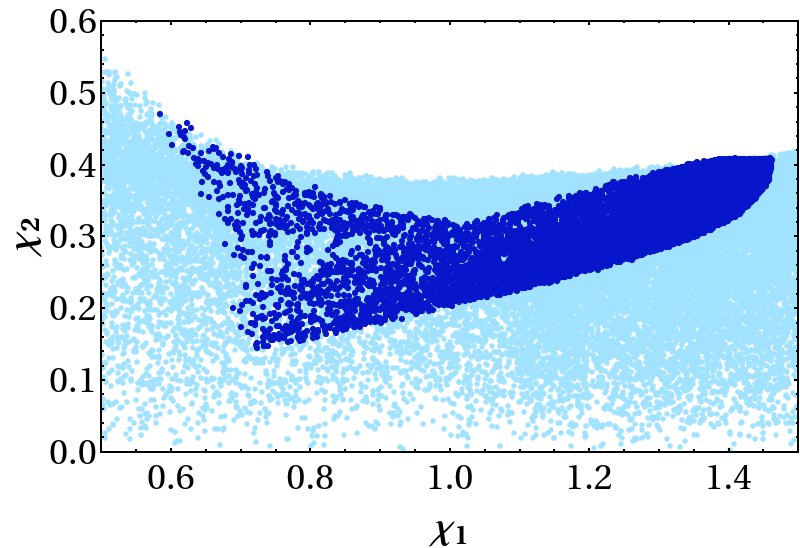}~~~~~~
			\includegraphics[width=0.5\linewidth]{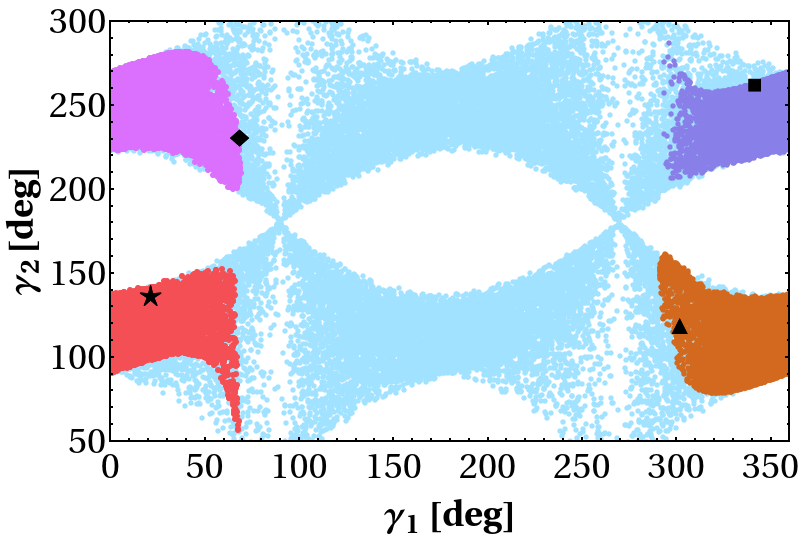}
			$$
			\caption{Allowed parameter spaces of $\chi_1$-$\chi_2$ (left panel) and  
					$\gamma_1$-$\gamma_2$ (right panel) using 3$\sigma$ ranges of neutrino oscillation parameters~\cite{Esteban:2020cvm}. The light blue dots in both the panels correspond to 3$\sigma$ allowed values for the mixing angles while the darker patches in each panel further satisfy constraints coming from the mass-squared differences, their ratio and sum of absolute masses. $\bigstar, \blacktriangle, \blacksquare, \blacklozenge$ marks of the right panel are indicative of four benchmark points (BP) used in Section \ref{sec:lepto}.}
			\label{fig:1}
		\end{figure}
		Equipped with all these, we provide a range of the allowed parameter space of our model in Fig. \ref{fig:1}. In the left panel, we first indicate the correlation between two of the parameters $\chi_1-\chi_2$ while the same for $\gamma_1-\gamma_2$ is shown in the right panel, indicated by the light blue points. The corresponding values of the parameters (light blue points) 
		satisfy the 3$\sigma$ allowed ranges of the lepton mixing angles, $\theta_{13}, \theta_{12}, \theta_{23}$. In obtaining these points, we varied parameters within a large range. For example, $\chi_{1,2}$ are varied from 0 to 2 while $\gamma_{1,2}$ are considered within their full range: 0-$360^{\circ}$. Once we also incorporate the constraints following from the mass-squared differences as well as the one on the sum of the light neutrino masses, the entire allowed parameter space is reduced to a smaller region indicated by the dark blue patch on the left panel (in $\chi_1-\chi_2$ plane) and four cornered patches (red, magenta, brown and purple) on the right panel (in $\gamma_1-\gamma_2$ plane).   
		
		From Fig. \ref{fig:1}, we find $0.584 \lesssim \chi_1  \lesssim 1.462$ whereas the ratio of the magnitudes of the antisymmetric contribution to the diagonal one (in view of Eq. (\ref{Ynudef}})) falls in a range: $0.470  \gtrsim \chi_2 \gtrsim 0.145$. Turning into the right panel, we find that $\gamma_1$ and $\gamma_2$ both are pushed toward four cornered regions represented by red, magenta, brown and purple patches respectively. Here we find that for $  0^{\circ}\leq \gamma_1 \leq 69^{\circ}$ the allowed regions for $\gamma_2$ are ($57^{\circ}-152^{\circ}$) and ($200^{\circ}-282^{\circ})$. Whereas for 
	$  291^{\circ}\leq \gamma_1 \leq 360^{\circ}$, the allowed regions for $\gamma_2$ are limited within  ($78^{\circ}-161^{\circ}$) and ($206^{\circ}-287^{\circ})$.
	Here we also note that,  in the right panel of Fig. \ref{fig:1},    $\bigstar , \blacktriangle, \blacksquare$ and $ \blacklozenge$ represent  four unique benchmark points in the parameter space $\{\chi_1, \chi_2, \gamma_1, \gamma_2\}$ given by BP1 = ($1.37, 0. 399, $ $21.53^{\circ}, 135.59^{\circ}$), BP2 =  ($0.978,0.235, $ $301.81^{\circ}, 119.1^{\circ}$), BP3 = ($ 1.417,0.372, $ $341.6^{\circ}, 260.83^{\circ}$) and BP4 = ($0.707, 0.209, $ $68.62^{\circ},~ 231.15^{\circ}$). 
	It is important to note that so far 
	the analysis presented here is applicable only for normal hierarchy of light neutrino mass. In the present setup, due to the special flavor structure of the model an inverted hierarchy of light neutrino mass spectrum however can not be accommodated. This is an interesting prediction that will undergo tests in several ongoing and near-future experiments. 
	
	\subsection{Implications for light neutrino masses and low energy phase}
	
	From the previous part of the analysis, we have an understanding on the allowed regions for the $\chi_1, \chi_2, \gamma_1$ and $\gamma_2$ which satisfy all  the constrains in the form of mass square differences, mixing angles and sum of the light neutrino masses. Hence, we are now in a position to study the implications  of this allowed parameter space toward the predictions involving sum of the light neutrino masses, and phases. We already have correlation between the Dirac CP phase $\delta$ and the atmospheric mixing angle $\theta_{23}$ as seen from Eqs. (\ref{col1}) and (\ref{col2}), both of which are functions of $\chi_{1,2}, \gamma_{1,2}$ as evident from Eqs. (\ref{theta}), (\ref{psi}). 
	\begin{figure}[h]
		\begin{subfigure}{.45\textwidth}
			\centering
			\includegraphics[width=1\textwidth]{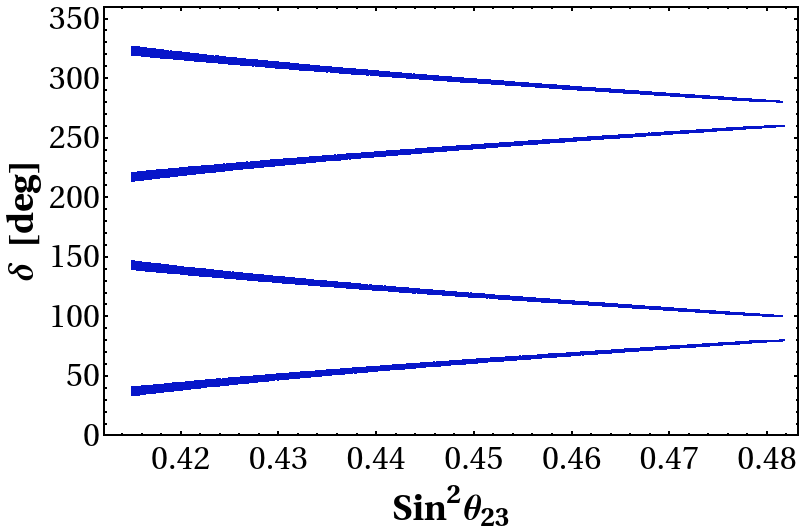}
			\caption{ } 
			\label{fig:2a}
		\end{subfigure}
		\begin{subfigure}{.45\textwidth}
			\centering
			\includegraphics[width=1\textwidth]{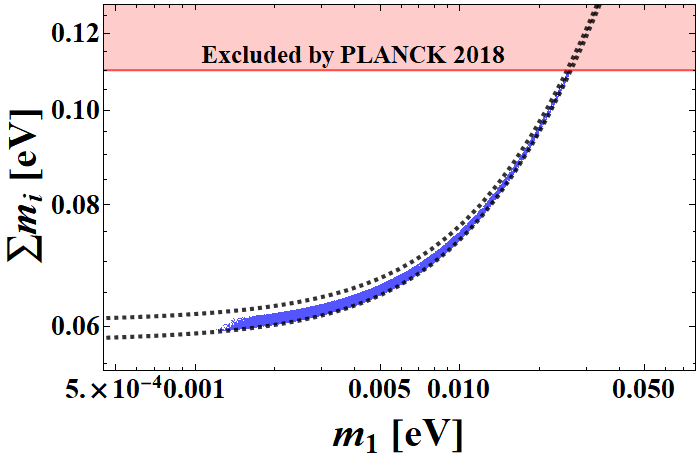}
			\caption{} 
			\label{fig:2b}
		\end{subfigure}
		\caption{Correlations within $\delta-\theta_{23}$  (left panel) and $\sum_i m_i-m_1$ (right panel) are presented while allowed ranges for $\chi_1$, $\chi_2$, $\gamma_1$ and $\gamma_2$ are used from Fig. \ref{fig:1}.}
	\end{figure}
	In Fig. \ref{fig:2a}, we have plotted this correlation in $\delta-\theta_{23}$ plane where only the allowed 
	set of points for $\chi_1, \chi_2, \gamma_1$ and $\gamma_2$ are employed (as in Fig. \ref{fig:1}). The model seems to predict 
	$\delta$ to be in the range $33^{\circ}(213^{\circ})\lesssim \delta \lesssim 80^{\circ}(260^{\circ})$  and  $100^{\circ}(280^{\circ}) \lesssim \delta \lesssim 147^{\circ}(327^{\circ})$ which 
	correspond to the atmospheric mixing angle $\theta_{23}$ in the lower octant. 
	Similarly, in Fig. \ref{fig:2a}, we use Eqs. (\ref{m1}), (\ref{m2}), (\ref{m3}) along with Eq. (\ref{prefac}) to  indicate the predictions related to the sum of the light neutrino masses against $m_1$, the lightest neutrino mass, indicated by the blue patch.
	The region between the black dotted lines represents 3$\sigma$ allowed range for $\sum_i m_i$ and the blue patch within it represents the predicted region in our framework. The red shaded region with  $\sum_{i} m_i \leq 0.11$ eV is  disallowed by cosmological observation mentioned earlier.
	This plot shows that the lightest neutrino mass is $\mathcal{O}(10^{-3})$ eV whereas the sum of the light neutrino masses is around $\mathcal{O}(0.06)$ eV. On top of this, the present set up excludes the possibility of having maximum CP violation ($\delta= 90^{\circ}/ 270^{\circ}$) and at the same time favors $\theta_{23}$ to be below maximal mixing, $i.e.~ \theta_{23} < 45 ^{\circ}$. These are the salient features of our proposal.

	\subsection{Neutrinoless Double beta decay}
	
	\begin{figure}[h]
		
		$$
		\includegraphics[width=.7\linewidth]{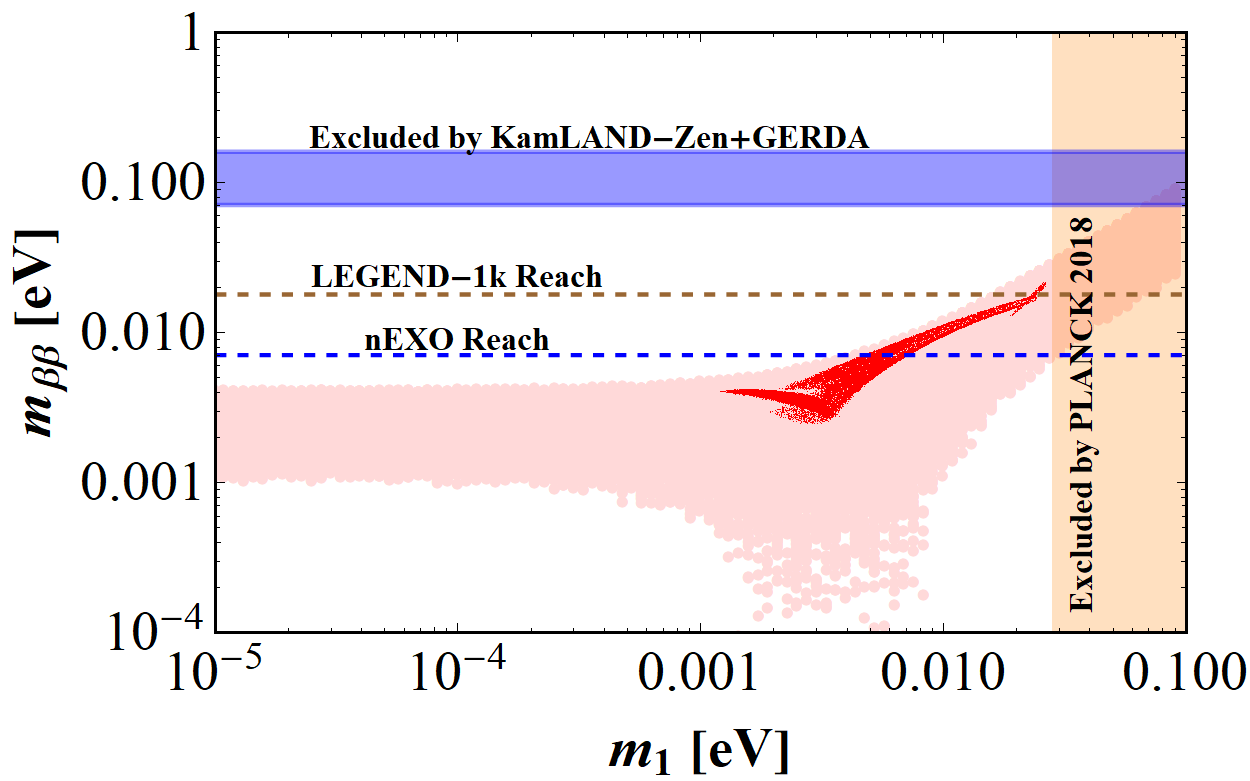}
		$$
		\caption{Correlation between $m_{\beta\beta}$ and lightest neutrino mass $m_1$ (for NH) with allowed ranges for $\chi_1$, $\chi_2$, $\gamma_1$ and $\gamma_2$ obtained from Fig. \ref{fig:1}. Here  the light blue shaded region represents the combined upper limit  of GERDA and KamLAND-Zen experiments whereas the brown and blue dashed lines stand for future sensitivities of the LEGEND and nEXO experiments respectively. }
		\label{fig:4}
	\end{figure}
	
	It is pertinent to also shed light on the effective neutrino mass parameter, $m_{\beta\beta}$, involved in the half life of  neutrinoless double beta decay in our set-up, which is given by~\cite{Tanabashi:2018oca}
	\begin{align}\label{betad}
	m_{\beta\beta} = |m_1 c_{12}^2 c_{13}^2 + m_2 s_{12}^2 c_{13}^2 e^{i\alpha_{21}}+m_3 s_{13}^2 e^{i(\alpha_{31}-2\delta)}|.
	\end{align}
	Note that for the normal hierarchy of light neutrino masses, one can write  $m_2= \sqrt{(m_1^2+\Delta m_{21}^2)}$, and $m_3= \sqrt{(m_1^2+\Delta m_{31}^2)}$. Recall also that we have already elaborated on our finding for lightest neutrino masses $m_{1}$ (see Fig. \ref{fig:2b}), and $\delta$ (see Fig. \ref{fig:2a}) in the last subsections corresponding to the allowed parameter space of $\{ \chi_1, \chi_2, \gamma_1, \gamma_2\}$ from Fig. \ref{fig:1}. Using the same, we could also 
	estimate the respective allowed ranges of Majorana phases $\alpha_{21}$ and $\alpha_{31}$ via Eq. (\ref{B}) (as $\alpha_{21} = \beta_{21}$ and $\alpha_{31} = \beta_{31}$) and in turn we can evaluate $m_{\beta\beta}$ as function of $m_1$ (substituting $m_2$ and $m_3$ in Eq. (\ref{betad})). With the allowed ranges for $\chi_1$, $\chi_2$, $\gamma_1$ and $\gamma_2$  satisfying all the neutrino data inclusive of the cosmological mass bounds ($i.e.$ corresponding to the dark blue patch of left panel, and four cornered patches of right panel of Fig. \ref{fig:1}), we therefore plot $m_{\beta\beta}$ as a function of lightest neutrino masses $m_{1}$ for normal hierarchy as presented in Fig. \ref{fig:4} by the red patch. The background light red patch indicates 
	the allowed region in general when mixing angles, mass squared differences along with $\delta$ are allowed to vary within their 3$\sigma$ range. Hence from this $m_{\beta\beta}$ vs $m_{1}$ plot (red patch), we notice that for $m_1$ within the range (0.001-0.027) eV (allowed in our set-up as per Fig. \ref{fig:2b}), the effective mass parameter is predicted to be: $0.002 \lesssim m_{\beta\beta} \lesssim 0.021$ eV. This prediction lies well within the limits on $m_{\beta\beta}$ by combined analysis of GERDA and KamLAND-Zen experiments denoted by the light blue shade. The horizontal brown 
	and blue dashed lines stand for future sensitivity by the LEGEND and nEXO experiments.

	\subsection{Lepton flavor violation}
	Due to the existence of active-sterile neutrino mixing, the possibility of rare lepton flavor violating processes should arise in our framework. Out of all the processes, contribution to $\mu \to e \gamma$ is the most important one as it is significantly constrained. In the weak basis, $i.e.$ where charged and RHN  mass matrix is diagonal, the branching ratio of the same process can be written as~\cite{Ilakovac:1994kj, Tommasini:1995ii}:
	\begin{align}
	B(\mu \to e \gamma)= \frac{3 \alpha}{8 \pi}\Biggl\lvert \sum_i \mathbb{R}_{e i} \mathbb{R}^{\dagger}_{i \mu} \mathbb{F}\Big(\frac{M_i^2}{M_W^2}\Big)\Biggl\rvert^2,
	\end{align}
	where $\alpha= e^2/4\pi$ is the fine structure constant, $M_W$ stands for $W^{\pm}$ mass,  $\mathbb{R}= m_D M^{-1}_R$ is the mixing matrix representing active-sterile mixing, $M_i$ is the mass of RHN  mass eigenstates $N_i$ and $\mathbb{F}(x)=\frac{x(1-6x+3x^2+2x^3-6x^2 \ln{x})}{2(1-x)^4}$, with $x=M_{i}/M_W$. The current upper bound on the branching ratio of the $\mu \to e \gamma$  is found to be  ${\rm BR} (\mu \to e \gamma) \lesssim 4.2 \times 10^{-13}$ (at 90\% C.L.)\cite{Tanabashi:2018oca}. In our analysis,  with the allowed ranges for $\chi_1$, $\chi_2$, $\gamma_1$ and $\gamma_2$ (obtained from Fig. \ref{fig:1}) and $M_i$ in the TeV scale, the contribution towards the branching ratio for $\mu \to e \gamma$ turns out to be insignificant ($\mathcal{O}(10^{-35})$) compared to the experimental limit. \\
	
	\section{Leptogenesis}\label{sec:lepto}
	
	The presence of RHNs in the seesaw realization of light neutrino mass provides an opportunity to study leptogenesis from the CP-violating out-of-equilibrium decay of RHNs into lepton and Higgs doublets in the early universe\cite{Fukugita:1986hr,Datta:2021elq,Bhattacharya:2021jli}. The lepton asymmetry created is expected to be converted to a baryon asymmetry via the sphaleron process\cite{Khlebnikov:1988sr,Arnold:1987zg}. In the previous part of our analysis, we have found that the phenomenology of the neutrino sector is mainly dictated by four parameters $i.e$ $\chi_1, \chi_2, \gamma_1$, and $\gamma_2$ which in turn determine most of the observables in the neutrino sector. However we also notice the presence of the prefactor $|f_3^{\nu^2}|v^2/M$ associated to the light neutrino mass eigenvalues as given in Eq. (\ref{m1})-(\ref{m3}). 
	Using Eq. (\ref{prefac}), though this prefactor can be evaluated, we can't have specific estimate for the degenerate mass of the RHNs ($M$) as $f_3^{\nu}$ remains undetermined. To have a more concrete picture, we provide a plot for $|f_3^{\nu^2}|v^2/M$ against one of the parameters, $\chi_1$, in Fig. \ref{fig:5} obtained using the correlation with other parameters fixed by neutrino oscillation and cosmological data. Hence barring the ambiguity in determining $f_3^{\nu}$ apart from a  conservative limit $|f_3^{\nu}| < \mathcal{O}(1)$, $M$ is seen to be anywhere from a very large value (say $10^{14-15}$ GeV) to a low one (say TeV). 
	\begin{figure}[h]
		$$
		\includegraphics[width=0.6\linewidth]{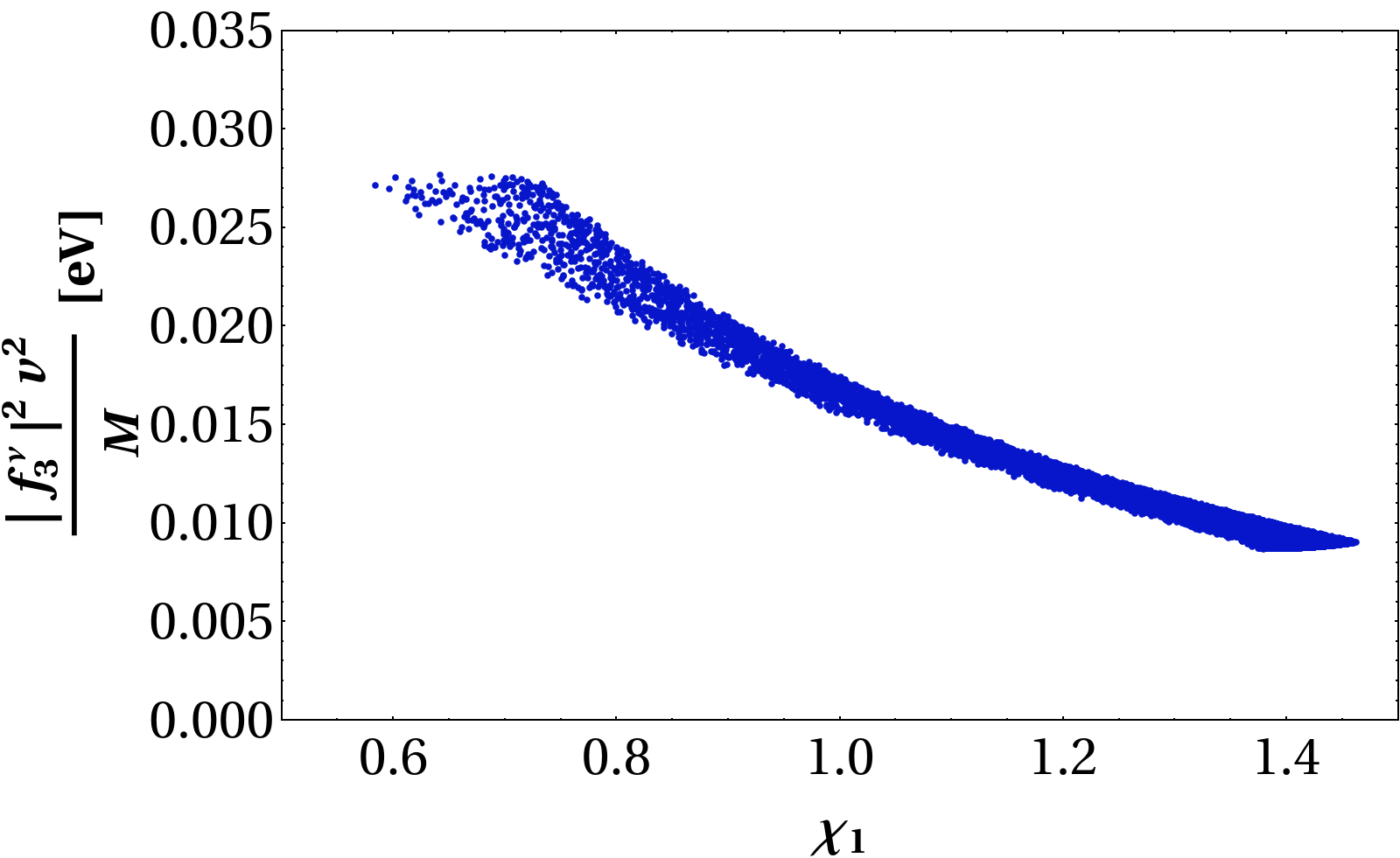}
		$$
		\caption{Correlation between $\frac{|{f^{\nu}_3}|^2 v^2}{M}$ and $\chi_1$ for NH.}
		\label{fig:5}
	\end{figure}	
	Furthermore, the RHNs are exactly degenerate in our framework. Hence unless we break this exact degeneracy, no 
	CP asymmetry can be generated \cite{Pilaftsis:1997jf} . Below we proceed to discuss leptogenesis mechanism in the present framework keeping in mind that we need to remove the exact degeneracy of RHN masses and study of flavored leptogenesis becomes essential (as $M$ can be below $10^{12}$ GeV).

	\subsection{Generation of mass splitting and CP asymmetry}
	
	The CP asymmetry parameter generated as a result of the interference between the tree and one loop level decay amplitudes of RHN $N_i$ decaying into a lepton doublet with specific flavor $l_{\alpha}$ and Higgs ($H$) is defined 
	by :
	\begin{align}
	\epsilon_{i}^{\alpha}= \frac{\Gamma(N_i \to \ell_{\alpha} H )-\Gamma(N_i \to \overline{\ell}_{\alpha} \bar{H} )}{\Gamma(N_i \to \ell_{\alpha} H )+\Gamma(N_i \to \overline{\ell}_{\alpha} \bar{H} )}.
	\end{align}
	Considering the exact mass degeneracy is lifted by some mechanism (will be discussed soon),
	the general expression for such asymmetry can be written as \cite{ Adhikary:2014qba,Pilaftsis:2003gt} :
	\begin{align}
	\epsilon_i^\alpha=&\frac{1}{8\pi \mathcal{H}_{ii}}\sum_{j\neq i} \text{Im}[\mathcal{H}_{ij} (\mathcal{Y^{\nu}}^{\dagger})_{i\alpha}(\mathcal{Y^{\nu}})_{\alpha j}] \Big[f(x_{ij}) + \frac{\sqrt{x_{ij}}(1-x_{ij})}{(1-x_{ij})^2+\frac{\mathcal{H}_{jj}^2}{64 \pi^2}}\Big]\nonumber\\
	& + \frac{1}{8\pi \mathcal{H}_{ii}}\sum_{j\neq i} \text{Im}[\mathcal{H}_{ji} (\mathcal{Y^{\nu}}^{\dagger})_{i\alpha}(\mathcal{Y^{\nu}})_{\alpha j}] \Big[\frac{(1-x_{ij})}{(1-x_{ij})^2+\frac{\mathcal{H}_{jj}^2}{64 \pi^2}}\Big]\label{eexp}, 
	\end{align}
	where $\mathcal{Y^{\nu}} ~(\equiv V^{\dagger} Y^{\nu}$ in our case, see Eq. (\ref{diagMD})) is the neutrino Yukawa matrix in charge lepton diagonal basis, $\mathcal{H}$ and the loop factor $f(x_{ij})$ are given by
	\begin{align}
	\mathcal{H}=& ~\mathcal{Y^{\nu}}^{ \dagger}\mathcal{Y^{\nu}}= ~Y^{\nu \dagger}Y^{\nu};\label{H}\\
	f(x_{ij})=&~\sqrt{x_{ij}}\Big[1-(1+x_{ij})\ln\left(\frac{1+x_{ij}}{x_ij}\right) \Big],
	\end{align}
	with $x_{ij}= \frac{M_j^2}{M_i^2}$ where $M_i$ are the masses of the RHNs after the degeneracy is removed. This  is applicable for both hierarchical as well as quasi-degenerate mass spectrum of RHNs \cite{Adhikary:2014qba}. For the hierarchical RHNs, one neglects  $\frac{\mathcal{H}_{jj}^2}{64\pi^2}$ compared to $(1-x_{ij})^2$ while the entire expression of Eq. ({\ref{eexp}}) can be used for quasi-degenerate case inclusive of resonance situation for which $(1-x_{ij})^2\simeq \frac{\mathcal{H}_{jj}^2}{64\pi^2}$ ~\cite{Pilaftsis:2003gt, Dev:2017wwc}. Below we discuss the mass splittings induced by the running of the heavy RHNs. 
	
	\subsubsection{Lifting the mass degeneracy}
	
	The exact mass degeneracy of heavy Majorana neutrinos is the result of the flavor symmetry imposed in our construction. To remove this degeneracy, here we adopt the renormalization group effects into consideration \cite{Casas:1999tp,GonzalezFelipe:2003fi}. Considering the discrete $A_4\times Z_3\times Z_2$ symmetry breaking scale close to the GUT scale $\sim \Lambda$ (the cut-off scale introduced in Eq. (\ref{eq:lag1})), we determine the running  
	of the RHN mass matrix $M_R$ and Dirac neutrino Yukawa matrix $\mathcal{Y}^\nu$ from GUT scale to seesaw scale $M$ (assuming $M < \Lambda$). 
	Using renormalisation group equations, the evolution of the RHN mass matrix $M$ (= diag($M_1, M_2, M_3$)) and Dirac neutrino Yukawa matrix $\mathcal{Y}^\nu$ (in charged lepton $Y^{\ell}$ diagonal basis) at one-loop can be written as~\cite{GonzalezFelipe:2003fi, Casas:1999tp,Chankowski:2001mx} 
	\begin{align}
	\label{RG1}
	\frac{d M_i}{dt}&=2 M_i\, \mathcal{H}_{ii}\,,\\
	\frac{d \mathcal{Y^{\nu}}}{dt}&=\left[ \{ {\mathcal{T}}- \frac{3}{4} g_1^2 - \frac{9}{4} g_2^2\}{\mathbb{I}}_{3}
	-\frac{3}{2} \left(Y^\ell Y^{\ell \dagger}-\mathcal{Y^{\nu}} \mathcal{Y^{\nu}}^{ \dagger} \right) \right]
	\mathcal{Y^{\nu}} + \mathcal{Y^{\nu}} R\,,
	\label{RG2}
	\end{align}
	with 
	\begin{align}
	{\mathcal {T}} = 3 \text{Tr}(Y_u Y_u^\dagger)+3 \text{Tr}(Y_d
	Y_d^\dagger)+\text{Tr}(Y^\ell Y^{\ell \dagger})+\text{Tr}(\mathcal{Y^{\nu}} \mathcal{Y^{\nu}}^{\dagger}),
	\end{align}
	where $Y_{u,d}$ are the up-quark and down-quark Yukawa matrices respectively, $g_{1,2}$ are the
	gauge couplings and $\mathbb{I}_3$ is the identity matrix of order $3 \times 3$. Here the matrix $R$ is anti-hermitian defined by~\cite{GonzalezFelipe:2003fi}
	\begin{align} 
	R_{11}&= R_{22}=R_{33}=0, \quad R_{ji} = -\; R_{ij}^\ast\,( i\neq j), \nonumber\\
	R_{ij}&=\frac{2+\delta_{ij}}{\delta_{ij}}\, \text{Re}\,(\mathcal{H}_{ij})+
	i\frac{\delta_{ij}}{2+\delta_{ij}}\, \text{Im}\,(\mathcal{H}_{ij})\,,
	\label{R}
	\end{align}
	$\delta_{ij} = \frac{M_j}{M_i}-1 $ is the degeneracy parameter for the RHN masses and $t = \frac{1}{16\pi^2}\ln \left(\frac{\Lambda}{M}\right)$.
	
	Now as the RHNs are exactly degenerate at scale $\Lambda$, the right hand side (first term) of Eq. (\ref{R}) becomes singular unless we impose $\text{Re}(\mathcal{H}_{ij})=0$. Note that, in our construction, $\mathcal{H}_{12}$ and $\mathcal{H}_{23}$ are already zero due to the flavor symmetry imposed. Hence the above condition should 
	be exercised only to realize $\rm{Re}(\mathcal{H}_{13}) =0$ in our case which can be materialized if 
	we choose to use $\mathcal{\tilde{Y}^{\nu}}$, obtained by performing an orthogonal rotation (by a matrix $O$ say) 
	on Dirac Yukawa matrix $\mathcal{Y^{\nu}}$ as,
	\begin{align}
	\mathcal{\tilde{Y}^{\nu}}=\mathcal{Y^{\nu}}O\,,{~\rm with} \quad O = \left(%
	\begin{array}{ccc}
	\cos \Theta &0 & \sin \Theta \\
	0& 1 &0 \\
	-\sin \Theta & 0 & \cos \Theta \\
	\end{array}%
	\right)\,,
	\end{align}
	having the rotation angle $\Theta$ determined by the relation
	\begin{align} \label{tan2theta}
	\tan 2 {\Theta} = \frac{2\text{Re}\,(\mathcal{H}_{13})}{\mathcal{H}_{33}-\mathcal{H}_{11}}\,=\frac{-\cos{\gamma_1}}{ \chi_2 \cos{(\gamma_1-\gamma_2)}}.
	\end{align}
	In obtaining the rightmost expression above, we employ Eqs. (\ref{Ynudef}), (\ref{chi1}), (\ref{chi2}) in Eq. (\ref{H}). This flexibility in using $\mathcal{\tilde{Y}^{\nu}}$ prevails due the following reason. Note that, if we rotate the $\mathcal{Y^{\nu}}$ in this manner, the neutrino Yukawa Lagrangian gets modified to:
	\begin{align}
	\bar{\ell_L} \mathcal{Y^{\nu}} \tilde{H} N_R = \bar{\ell_L}\mathcal{\tilde{Y}^{\nu}} O^T \tilde{H}N_R.
	\end{align}
	We can now redefine $N_R$ by: $\tilde{N}_R= O^T N_R$ , $i.e.$ if we rotate RHN fields by $O^T$, RHN mass term 
	will not change as $\overline{N_R^C} M_R N_R = \overline{\tilde{N}_R^C} M_R \tilde{N}_R$  due to  the orthogonal property of $O$  matrix. 
	
	The Eqs. (\ref{RG1}) and (\ref{RG2}) can now be rewritten in terms of $\tilde{\mathcal{H}} = O^T \mathcal{H} O$ 
	and $\mathcal{\tilde{Y}^{\nu}}$ by using the above relations. The form of $\tilde{\mathcal{H}}$ can be obtained by
	\begin{align} \label{Hnurotated}
	\tilde{\mathcal{H}}=\mathcal{\tilde{Y}^{\nu\dagger}}\mathcal{\tilde{Y}^{\nu}}=O^T \mathcal{H} O =\left(%
	\begin{array}{ccc}
	\mathcal{H}_{11}-\Delta &0 & i\, \text{Im}\,(\mathcal{H}_{13}) \\
	0 & \mathcal{H}_{22} & 0 \\
	-i\, \text{Im}\,(\mathcal{H}_{13}) &0 &  \mathcal{H}_{33}+\Delta \\
	\end{array}%
	\right)\,,
	\end{align} 
	\begin{figure}[t]
		\centering
		$$
		\includegraphics[width=0.25\linewidth]{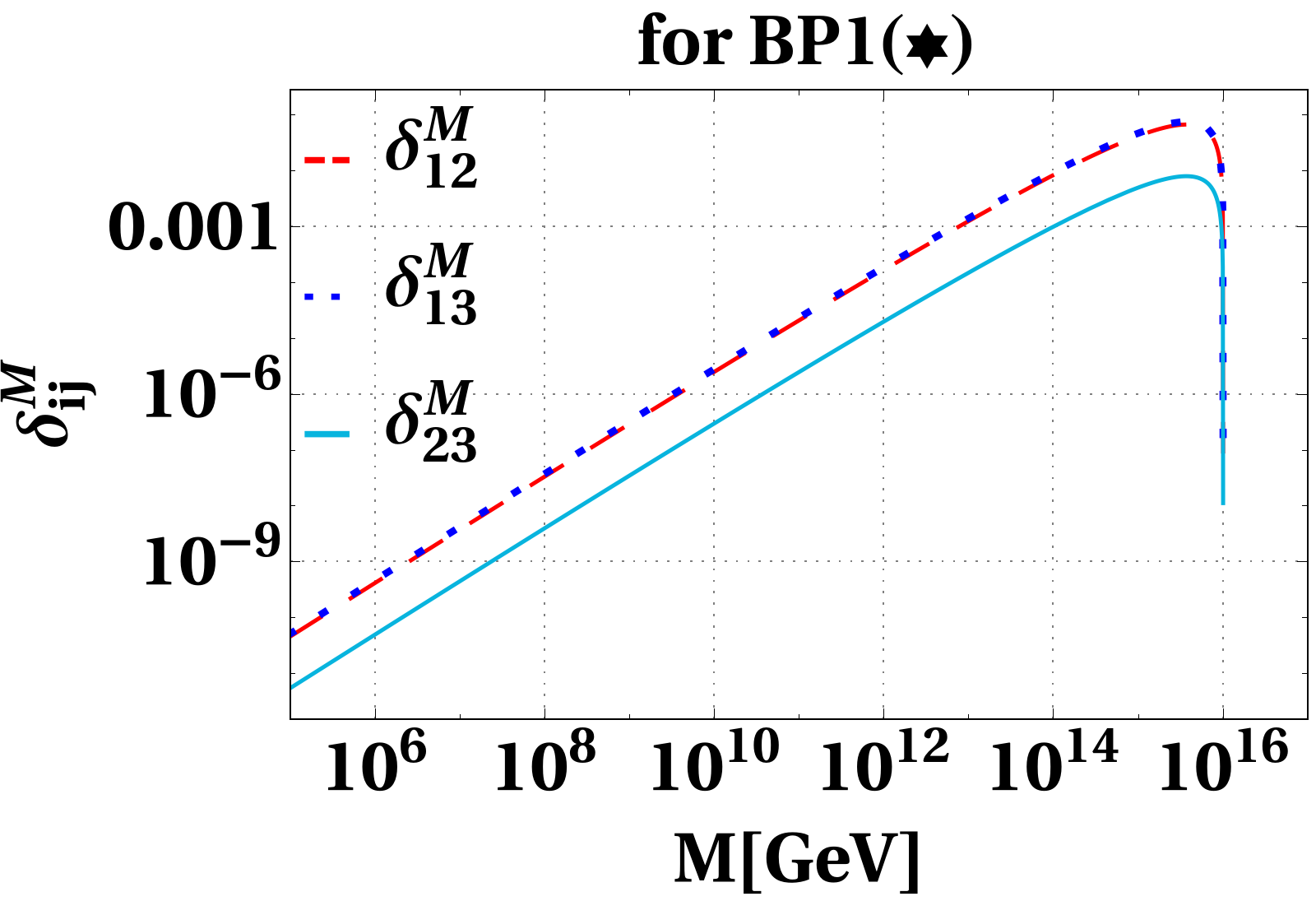}
		\includegraphics[width=0.25\linewidth]{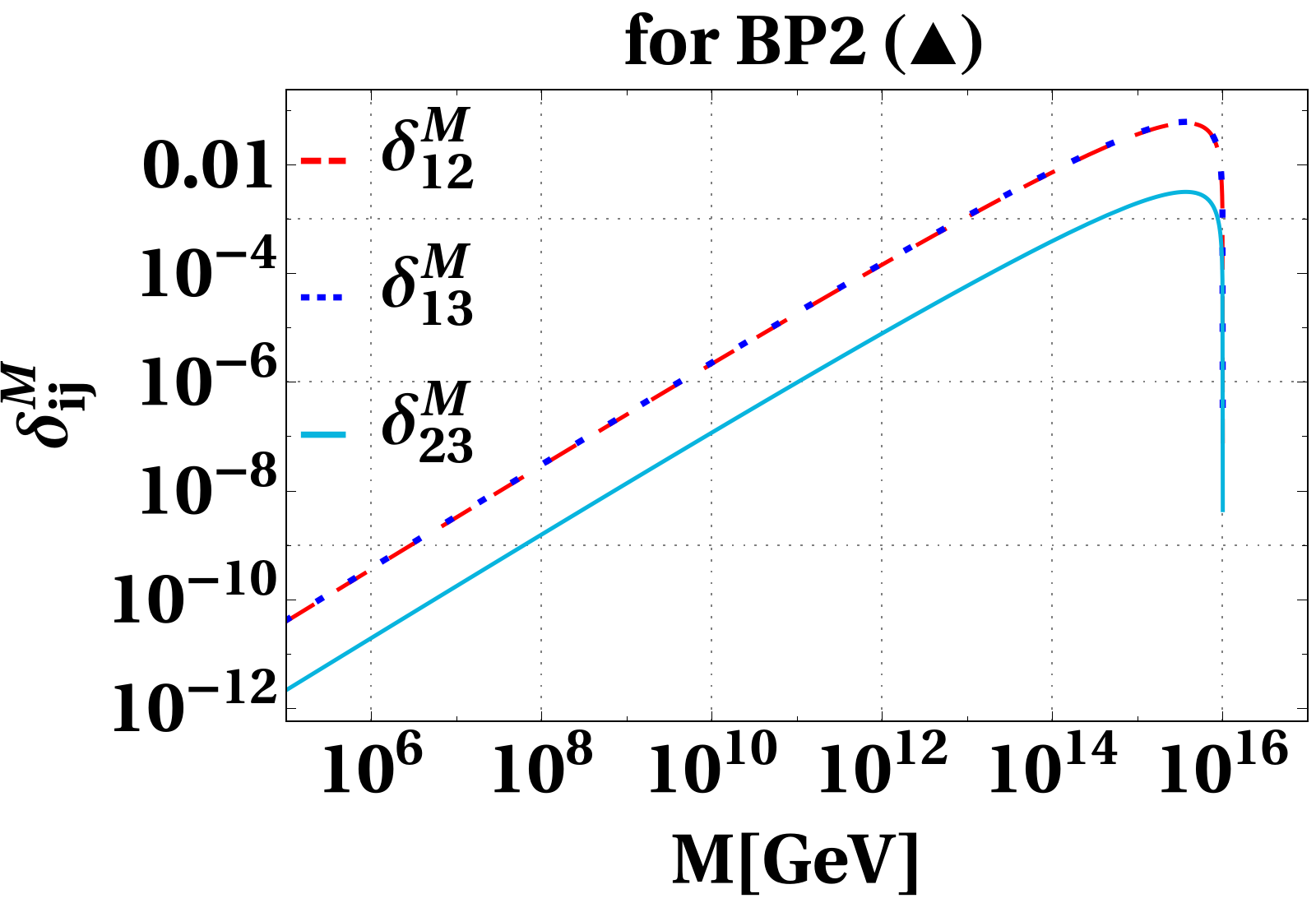}
		\includegraphics[width=0.25\linewidth]{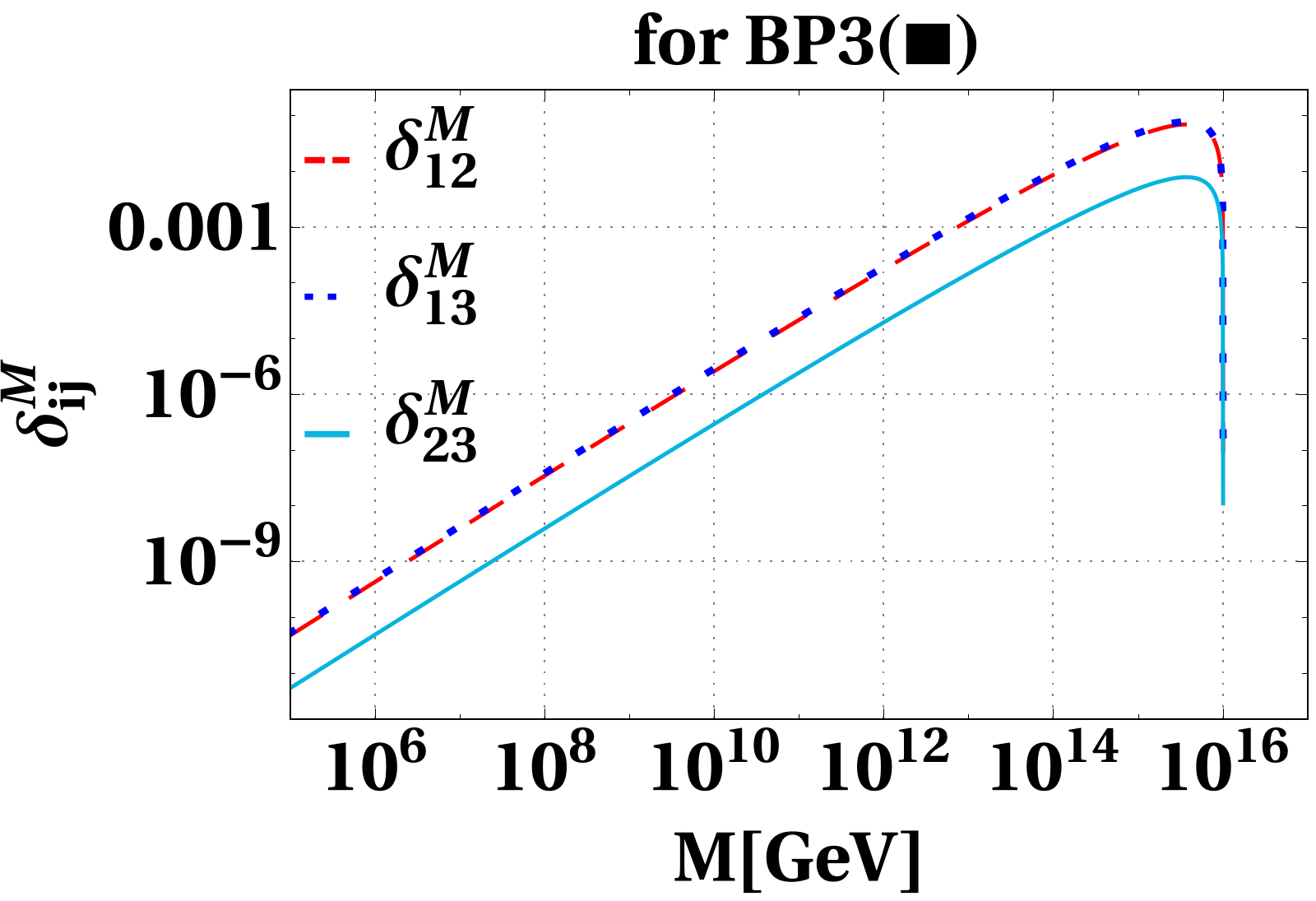}
		\includegraphics[width=0.25\linewidth]{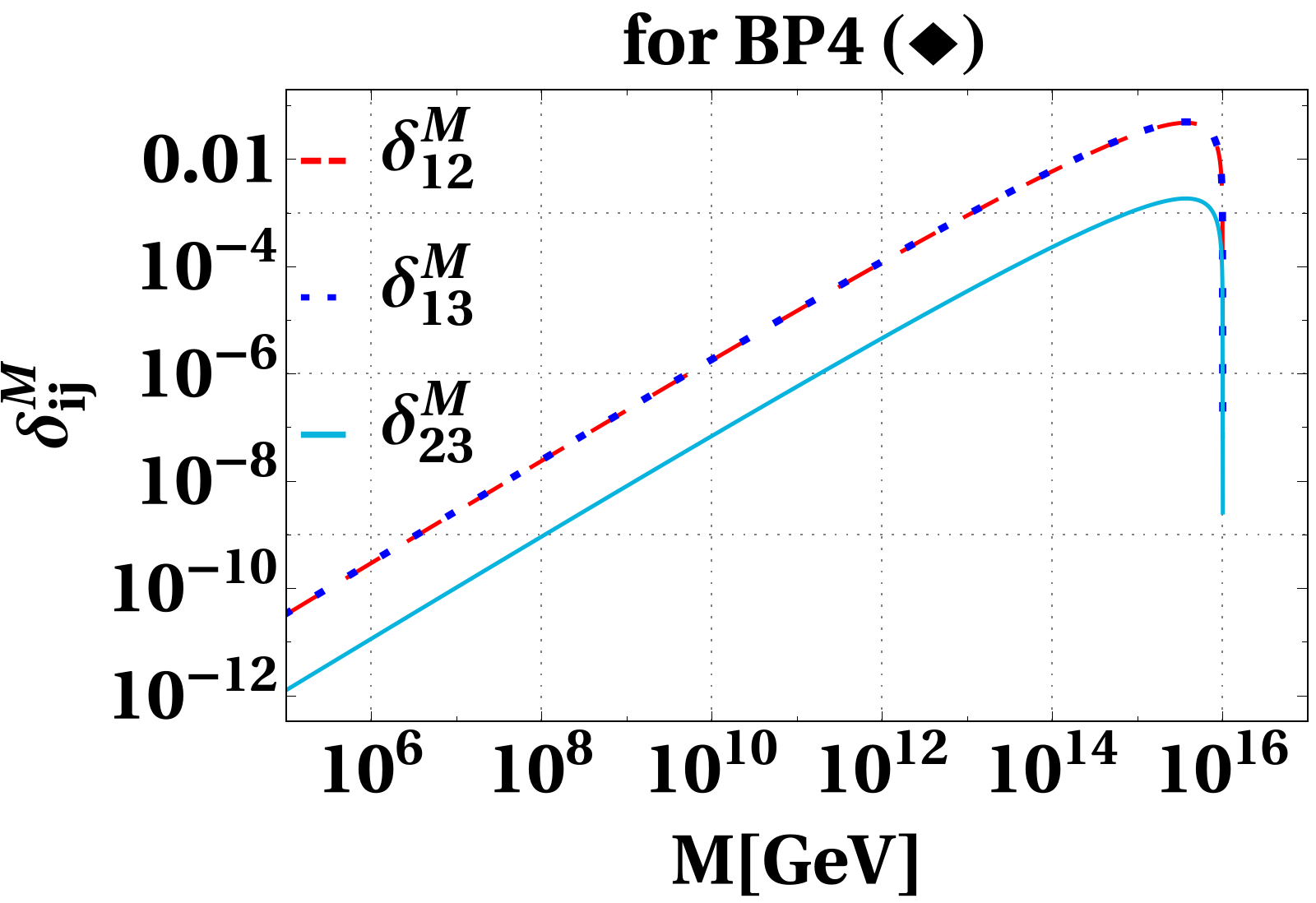}
		$$
		\caption{Variation of mass splitting $\delta^M_{ij}$ with respect to scale $M$ for the benchmark points BP1, BP2, BP3 and BP4 respectively. }
		\label{fig:del}
	\end{figure}
	where $\Delta \equiv \tan \Theta\, \text{Re}\,(\mathcal{H}_{13})$. As seen from the Eq. (\ref{RG1})
	(with right hand side written in terms of $\mathcal{\tilde{H}}$ now), we find that a mass splitting generated at a scale 
	$(M)$ as
	\begin{align}
	\delta_{ij}^M = 2 (\mathcal{\tilde{H}}_{ii}-\mathcal{\tilde{H}}_{jj})t\,,	
	\end{align}
	thanks to the effect of running. Using Eq. (\ref{RG2}), we also get a off-diagonal contribution ($\mathcal{H}^R_{ij}, i\neq j$) to $\mathcal{\tilde{H}}$ \cite{GonzalezFelipe:2003fi}, 
	\begin{align}
	\label{HR}
	{\mathcal{\tilde{H}}^M}_{ij} &= \mathcal{\tilde{H}}_{ij}+ \mathcal{H}^R_{ij}; ~~
	{\mathcal{H}}^R_{ij} \simeq 3 y_\tau^2\,\mathcal{\tilde{Y}^{\nu*}}_{3 i} \mathcal{\tilde{Y}^{\nu}}_{3 j}\,t; ~\quad(i\neq j)		\end{align}
	while ${\mathcal{\tilde{H}}^M}_{ii}  = \mathcal{\tilde{H}}_{ii}$. 
	As mentioned earlier, the seesaw scale $M$ remains undetermined even after applying neutrino mass and mixing constraints, we have shown in Fig. \ref{fig:del} how such splitting $\delta^M_{12}$ varies with the degenerate RHN mass $M$ due to running corresponding to benchmark points: BP1, BP2, BP3 and BP4 allowed by the neutrino data. We find that below $M \simeq 10^{12}$ GeV, $\delta_{ij}^M$ become smaller than $\mathcal{O}(10^{-4})$ implying that the masses of the three RHNs fall in the quasi-degenerate category \cite{Adhikary:2014qba}. Such a a small splitting, although crucial for generation of CP asymmetry, won't alter our findings of the neutrino section. Note that the estimated splitting does not correspond to the requirement of resonant leptogenesis. We are now in a position to evaluate the CP asymmetry generated at scale $M$, as discussed below.
	
	\subsubsection{Estimating CP asymmetry}
	
	\begin{figure}[h]
		\centering
		\includegraphics[width=1\linewidth]{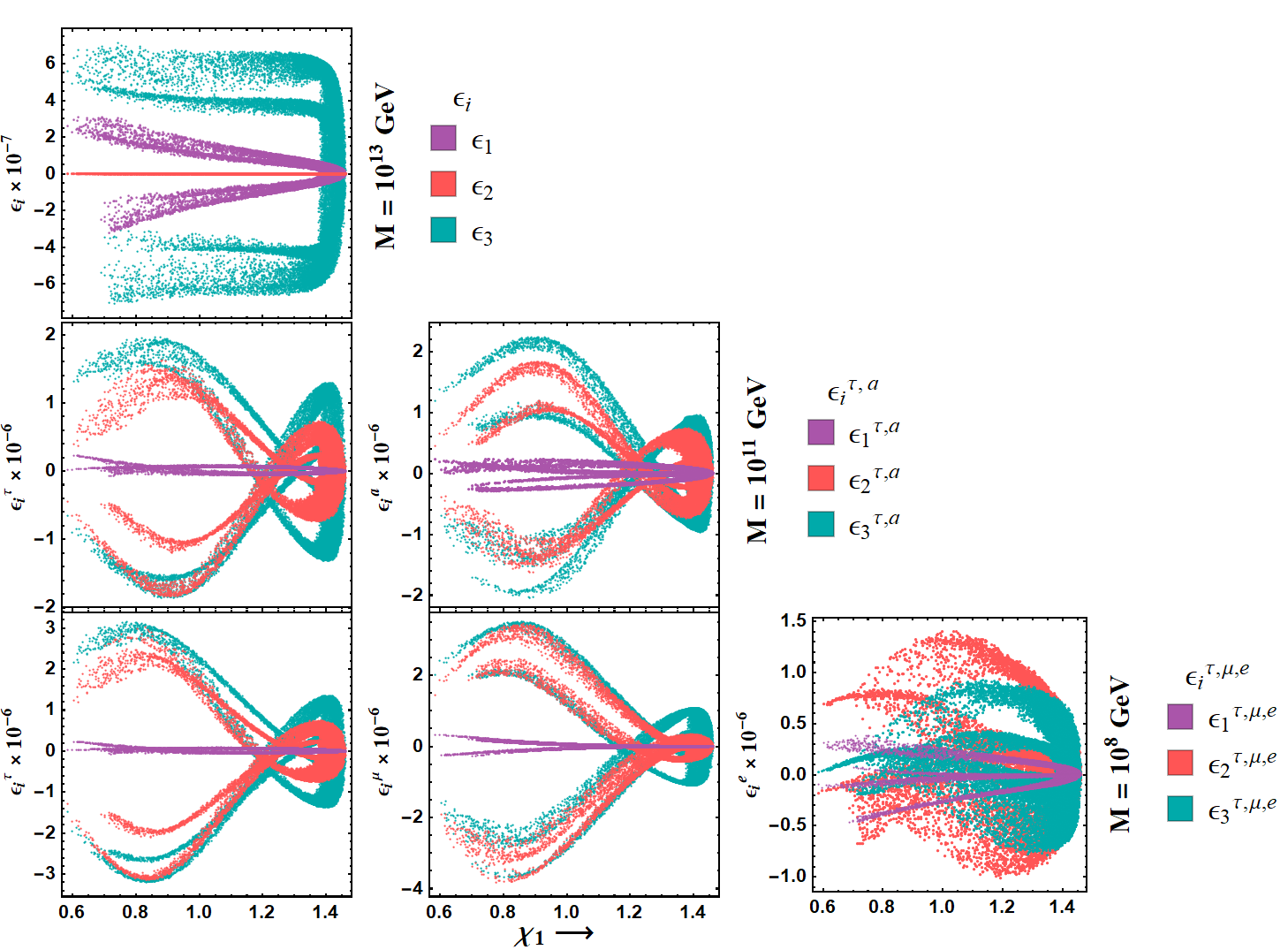}
		\caption{Variation of individual components of CP asymmetry with respect to model parameter $\chi_1$ for three different scales $M=10^{13}$ GeV (top most plot),  $M= 10^{11}$ GeV (plots from second row) and $M=10^8$ GeV (plots from third row).}
		\label{fig:10}
	\end{figure}
	Starting with exact degeneracy 	of RHN masses, we have shown that the running of involved parameters from a typical high scale to the scale of the heavy neutrino masses leads to a quasi-degenerate spectrum of RHNs. Hence we can now estimate the CP asymmetry created at a scale $M$ by using Eq. (\ref{eexp}) while replacing $\mathcal{H}$ by ${\tilde{\mathcal{H}}}^M$ and $\delta_{ij}$ by $\delta_{ij}^M$ in view of our discussion above. 
	Furthermore, it can be shown that maximum contribution to CP asymmetry comes from self energy diagram~\cite{Flanz:1996fb, Covi:1996wh, Pascoli:2006ci}. Therefore, the asymmetry expression of Eq. (\ref{eexp}) gets modified to  
	\begin{align}
	\epsilon^{\alpha}_i \simeq & ~  -\frac{1}{16\pi {\mathcal{\tilde{H}}^M}_{ii}}\sum_{j\neq i}\frac{\delta_{ij}^M}{(\delta_{ij}^M)^2+ \left(\frac{{\mathcal{\tilde{H}}^M}_{jj}}{16\pi}\right)^2}\,\Big\{\text{Im}[{\mathcal{\tilde{H}}^M}_{ij} \mathcal{\tilde{Y}^{\nu*}}_{\alpha i} \mathcal{\tilde{Y}^{\nu}}_{\alpha j}]+\text{Im}[{\mathcal{\tilde{H}}^M}_{ji} \mathcal{\tilde{Y}^{\nu*}}_{\alpha i} \mathcal{\tilde{Y}^{\nu}}_{\alpha j}]\Big\}\label{epfinal}.\,
	\end{align}
	Now, using Eqs. (\ref{Hnurotated}) to (\ref{HR}) and employing them in Eq. (\ref{epfinal}), we estimate for the cp asymmetry parameter for the heavy RHNs decaying into various flavors which will be useful to evaluate the final lepton asymmetry taking the flavor effects into account. Since all the entities of Eq. (\ref{epfinal}) are function of set of parameters $\{\chi_1, \chi_2, \gamma_1, \gamma_2\}$ and $M$, we can make use of the allowed parameter space from neutrino phenomenology (refer to Fig. \ref{fig:1}) and finally calculate the CP asymmetries produced from all three RHN decays ($i=1,2,3$) to different flavors of lepton doublets and Higgs. 
	
	For representation purpose, in Fig. \ref{fig:10}, we depict the variation of individual flavor components of CP asymmetry 
	with respect to $\chi_1$ at three different RHN mass scales: $M = 10^{13}$ (top panel), $10^{11}$ (middle panel), $10^8$ (bottom panel) GeV respectively. Since the flavor effects are known not to be important beyond $T \sim M \simeq 10^{12}$ GeV, we estimate asymmetries produced by individual RHNs only for top panel. It is found that maximum asymmetry falls 
	in the ballpark of $|\epsilon_{i=1,3}|_{\rm{max}} \sim 6 \times 10^{-7}$ whereas $(|\epsilon_2|)_{\rm{max}}$ remains subdominant. 
	At $ T = 10^{11}$ GeV (and above $10^8$ GeV), tau Yukawa comes to equilibrium, so effectively the scenario with $M = 10^{11}$ GeV becomes a two flavor scenario ($\tau$ and another orthogonal direction, say $a$) and the corresponding CP asymmetries are marked by: $\{\epsilon_{i}^{\tau},\epsilon_{i}^{a=\mu+e}\}$. At this scale, $|\epsilon_{i=2,3}^{\tau,a}|_{max}\sim 2\times 10^{-6}$ (middle panel of Fig. \ref{fig:10}) and $|\epsilon_1^{\tau,a}|_{\rm{max}}$ becomes relatively small. 
	We also estimate CP asymmetry at $M= 10^8$ GeV(bottom panel of Fig. \ref{fig:10}). At this temperature (or scale), all Yukawa couplings are in equilibrium and hence contributions to CP asymmetries from all the three flavors, $\{\epsilon_{i}^{e},\epsilon_{i}^{\mu},\epsilon_{i}^{\tau}\}$, become important. We find $|\epsilon_{i=2,3}^{\tau}|_{max} \sim 3\times 10^{-6}$ and $|\epsilon_1^{\tau}|_{max} < |\epsilon_{i=2,3}^{\tau}|_{max}$. An analogous pattern is observed for $\epsilon_i^{\mu}$. CP asymmetry along electron flavor is shown in the third plot of the bottom panel of Fig. \ref{fig:10} and is found to be $|(\epsilon_{i=2,3}^{e})_{max}|\sim 1.5\times 10^{-6}$ , $|(\epsilon_1^e)_{max}|\sim 5 \times 10^{-7}$.
	With these various flavor dependent CP asymmetries, we can now proceed for evaluation of baryon asymmetry by solving the Boltzmann equations as illustrated below. 	
	
	\subsection{Solution of Boltzmann equation}
	
	It is worth mentioning that while estimating the final lepton asymmetry, one needs to take care of decays and inverse decays of heavy RHNs as well as various scattering processes. As stated earlier, we consider the contributions of all 
	three RHNs having $M_{i} \lesssim 10^{12}$ GeV. Hence flavor effects have to be considered \cite{Abada:2006fw} 
	as with the mass equivalent temperature regime $T\sim10^{12}$ GeV, decay rate of $\tau$ $(\Gamma_{\tau}\sim 5 
	\times 10^{-3} y_{\tau}^2 T)$ \cite{Cline:1993bd, Abada:2006ea} becomes comparable to the Hubble expansion rate. 
	Below this temperature, the relation becomes $\Gamma_{\tau} > H$ indicative of the start of equilibrium era for $\tau$ Yukawa interactions and $\tau$ lepton doublet becomes distinguishable. In a similar way, for the temperature regime $10^{8}$ GeV $\lesssim T \lesssim 10^{11}$ GeV, muon Yukawa interaction comes to equilibrium (and both $\mu$ and $\tau$ flavors of lepton doublets are distinguishable henceforth) and finally below $T \lesssim 10^8$ GeV, $e$ Yukawa interaction are in equilibrium.
	
	In our analysis, therefore,  we include these flavor effects into consideration while constructing the Boltzmann equations.  We work in a most general setup for leptogenesis, where all three RHNs are contributing to the asymmetry due to their quasi degenerate spectrum of masses. As standard, the produced lepton doublets from the RHN decay needs to be appropriately projected to flavor states in the three above mentioned temperature regimes differently where the related respective lepton asymmetries are characterized by the $C^{\ell}$ matrices ($C^{H}$ stands for that of Higgs) \cite{Nardi:2006fx,Nardi:2005hs}. For example, when only the $\tau$ Yukawa interaction is in equilibrium ($10^{11}$ GeV $ \lesssim T \lesssim10^{13}$ GeV), effectively the scenario becomes a two flavor case (as the flavor space is spanned by $\ell_{\tau}$ and another orthogonal direction) and so $C^l$ is a matrix of order $2 \times 2$. For a further smaller temperature, the situation comprises of three effective flavors and so $C^{\ell}$ is of $3 \times 3$. Below we write down the relevant Boltzmann equations to study the time evolution of the lepton-number asymmetries (for a system of three RHNs) as ~\cite{Nardi:2006fx, Nardi:2005hs,Asaka:2018hyk}

	\begin{align}
	s H z \frac{d Y_{N_i}}{dz}
	&=
	-
	\left	\{	\left( \frac{Y_{N_i}}{Y_{N_i}^{\rm eq}}	-	1	\right)
	(\gamma_{D_i} + 2 \gamma_{N^i_s}	+4\gamma_{N^i_t})
	\ +	\sum_{j \neq i}
	\left(	\frac{Y_{N_i}}{Y_{N_i}^{\rm eq}}\frac{Y_{N_j}}{Y_{N_j}^{\rm eq}}	-	1	\right)
	(\gamma_{N_i N_j}^{(1)}+\gamma_{N_i N_j}^{(2)})
	\right \} \,,\label{be1}\\
	s H z \frac{d Y_{\Delta_{\alpha}}}{dz}
	&=
	-\left	\{	\sum_{i}
	\left(	\frac{Y_{N_i}}{Y_{N_i}^{\rm eq}}	-1\right)
	\epsilon_{i}^{\alpha} \gamma_{D_i}
	-
	\sum_{\beta} \left[ \sum_{i}
	\left(	\frac{1}{2} \left(C_{\alpha \beta}^{\ell}	-	C_{\beta}^{H}\right)
	\gamma_{D_i}^{\alpha}	\right.\right.\right.
	\notag	\\	
	&\left. \left.
	+ \left(C_{\alpha \beta}^{\ell}\frac{Y_{N_i}}{Y_{N_i}^{\rm eq}}
	-	\frac{C_{\beta}^{H}}{2}\right)\gamma_{N^i_s}
	+ \left(	2C_{\alpha \beta}^{\ell}	-\frac{C_{\beta}^{H}}{2}
	\left(	1+\frac{Y_{N_i}}{Y_{N_i}^{\rm eq}} \right) \right) \gamma_{N^i_t}
	\right) \right.
	\notag\\	
	&\left.\left.
	+ \sum_{\gamma} \left(
	\left(	C_{\alpha \beta}^{\ell}	+C_{\gamma \beta}^{\ell}		-2C_{\beta}^{H}\right)
	\left(	\gamma_{N}^{(1)\alpha \gamma} +	\gamma_{N}^{(2)\alpha \gamma}\right)
	\sum_{i,j}	\left(	C_{\alpha \beta}^{\ell}	-C_{\gamma \beta}^{\ell}\right)
	\gamma_{N_{i} N_{j}}^{(1) \alpha \gamma}	\right)
	\right] \frac{Y_{\Delta_{\beta}}}{Y^{\rm eq}}
	\right \} \,\label{be2},
	\end{align}
	where $z=M_i/T$ and $\alpha=e, \mu,\tau$. In the above, $Y_{\Delta_{\alpha}(N_i)}= n_{\Delta_{\alpha} (N_i)}/s$ denotes the density of $\Delta_{\alpha} = \frac{B}{3}-L_\alpha$ (relevant heavy neutrino) with respect to the entropy $s,Y^{\rm eq}$'s are the respective number densities while in thermal equilibrium. Here, total decay rate density of $N_i$ 
	is given by 
	\begin{align}
	\gamma_{D_i}= \sum_{\alpha}[ \gamma(N_i\rightarrow \ell_\alpha + H)+ \gamma(N_i\rightarrow\bar{\ell_\alpha}+\bar{H})]= n_{N_i}^{eq} \frac{K_1(z)}{K_2(z)}\Gamma_i, \label{gd}
	\end{align}
	
	where $\Gamma_i$ is the total decay rate of $N_i$ at tree level and written as 
	\begin{align}
	\Gamma_{i}= \sum_{\alpha}[ \Gamma(N_i\rightarrow \ell_\alpha+ H)+ \Gamma(N_i\rightarrow\bar{\ell_\alpha}+\bar{H})], 
	\end{align}
	and $\gamma_{N^i_s}, \gamma_{N^i_t}$ (both are Higgs mediated scattering process with change in lepton number 
	$\Delta L = 1$), $\gamma_{N_{i} N_{j}}^{(1)}, \gamma_{N_{i} N_{j}}^{(2)} $ (both are neutrino pair annihilation process) are the reaction rate densities for the scattering processes: $[N_i+ \ell \leftrightarrow Q + \bar{U}]_s$, $[N_i+\bar{Q}\leftrightarrow \bar{\ell}+\bar{U}]_t+[ N_i+U\leftrightarrow \bar{\ell}+\bar{Q}]_t$, $[N_i+N_j\leftrightarrow \ell + \bar{\ell}]$ and   $[N_i+N_j\leftrightarrow H + \bar{H}]$ respectively\cite{Plumacher:1998ex,Plumacher:1996kc}. Here in Eq. (\ref{gd}), $K_1(z)$ and $K_{2}(z)$ are the modified Bessel functions.
	
	\begin{figure}[t]
		\begin{subfigure}{.50\textwidth}
			\centering
			\includegraphics[width=1\linewidth]{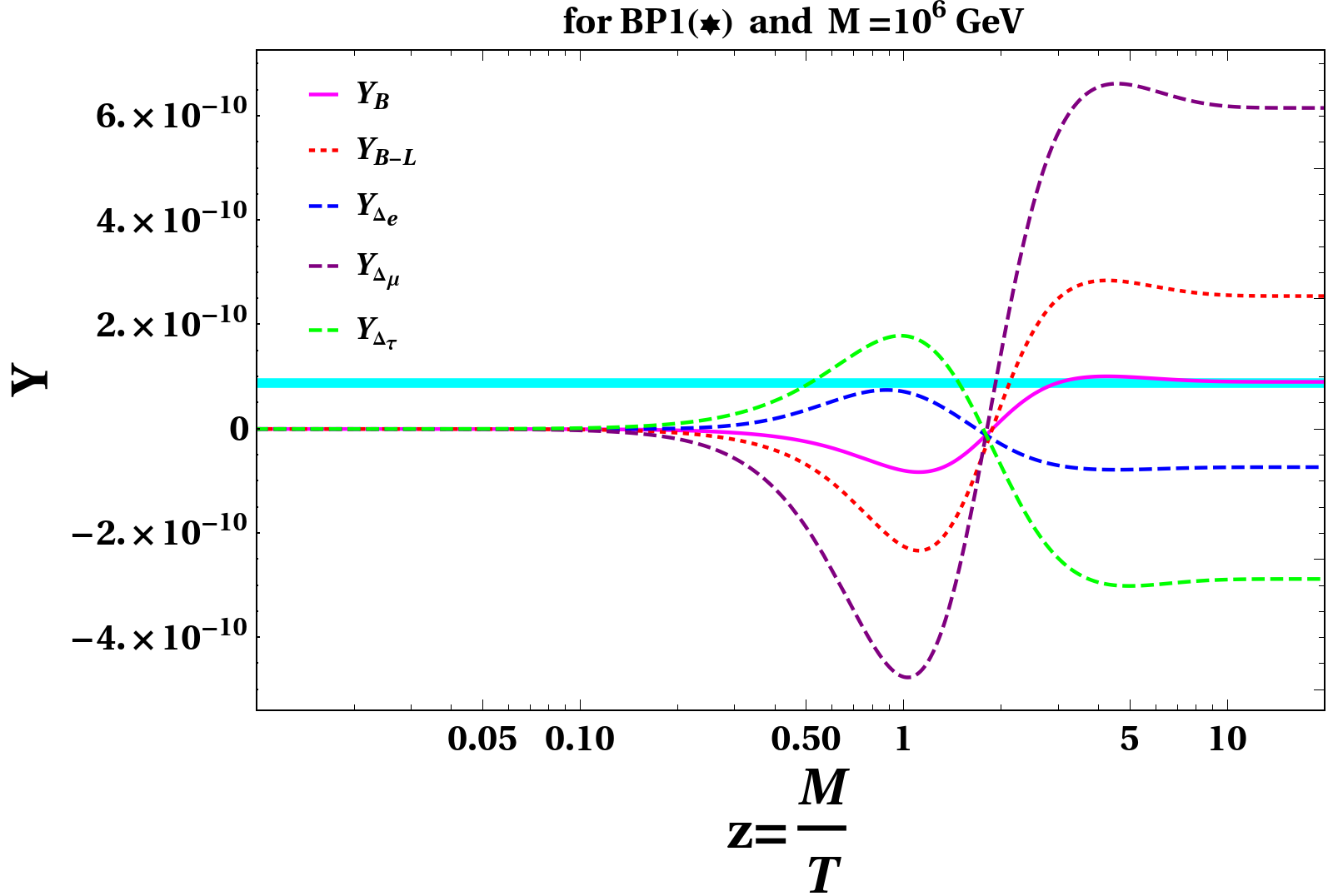}
			\caption{}
			\label{fig:8a}
		\end{subfigure}
		\begin{subfigure}{.50\textwidth}
			\centering
			\includegraphics[width=1\linewidth]{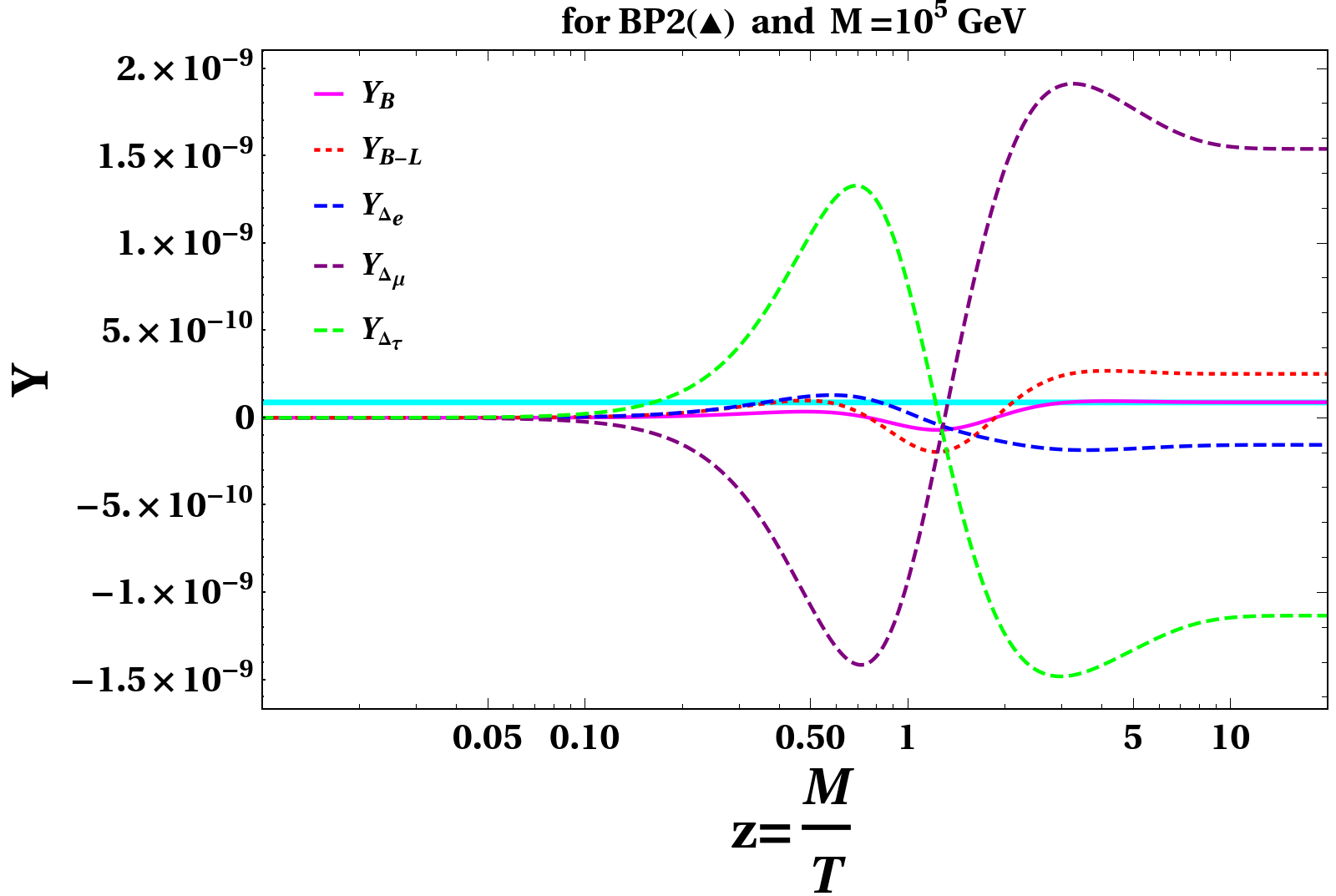}
			\caption{}
			\label{fig:8b}
		\end{subfigure}
		\begin{subfigure}{.50\textwidth}
			\centering
			\includegraphics[width=1\linewidth]{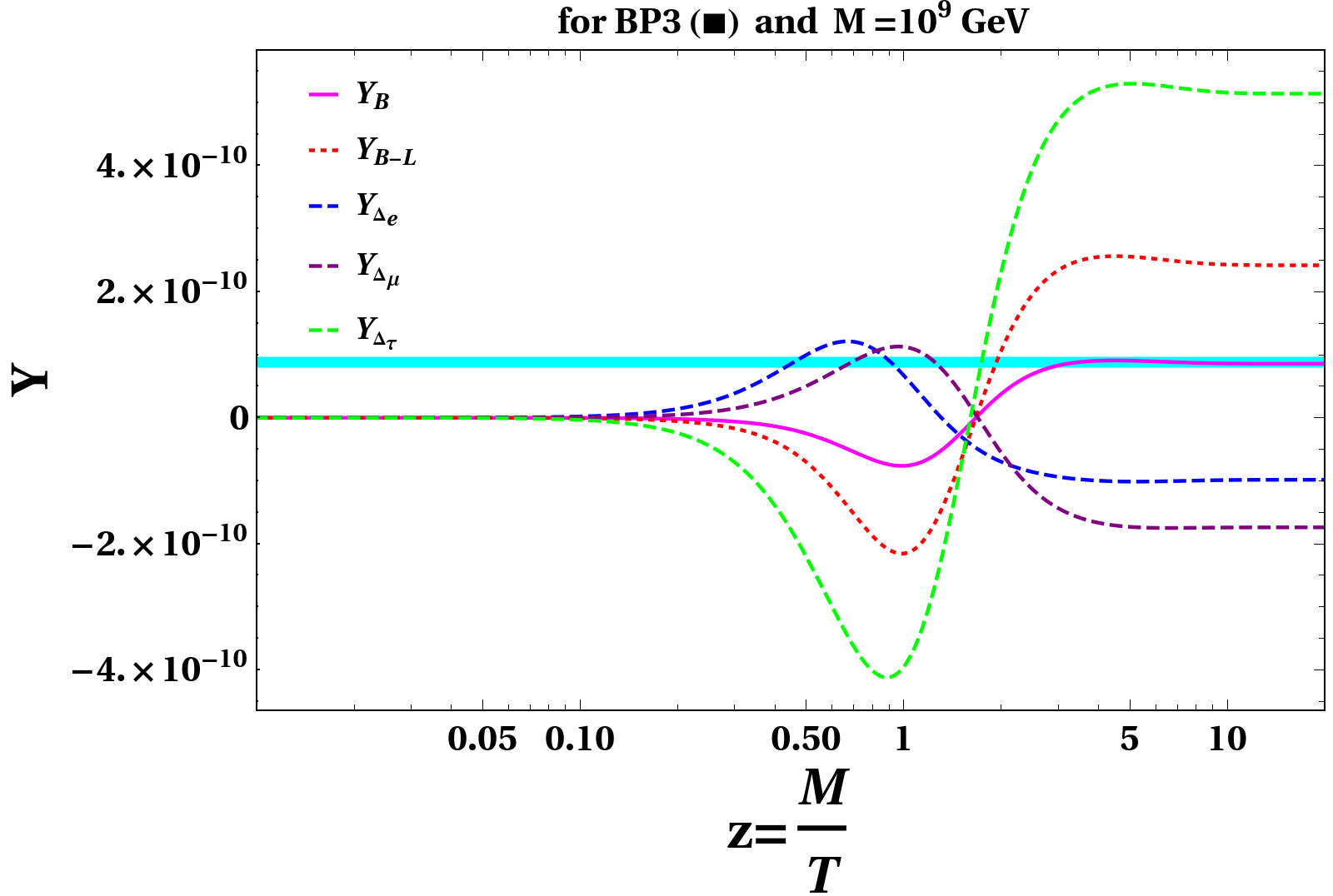}
			\caption{}
			\label{fig:8c}
		\end{subfigure}
		\begin{subfigure}{.50\textwidth}
			\centering
			\includegraphics[width=1\linewidth]{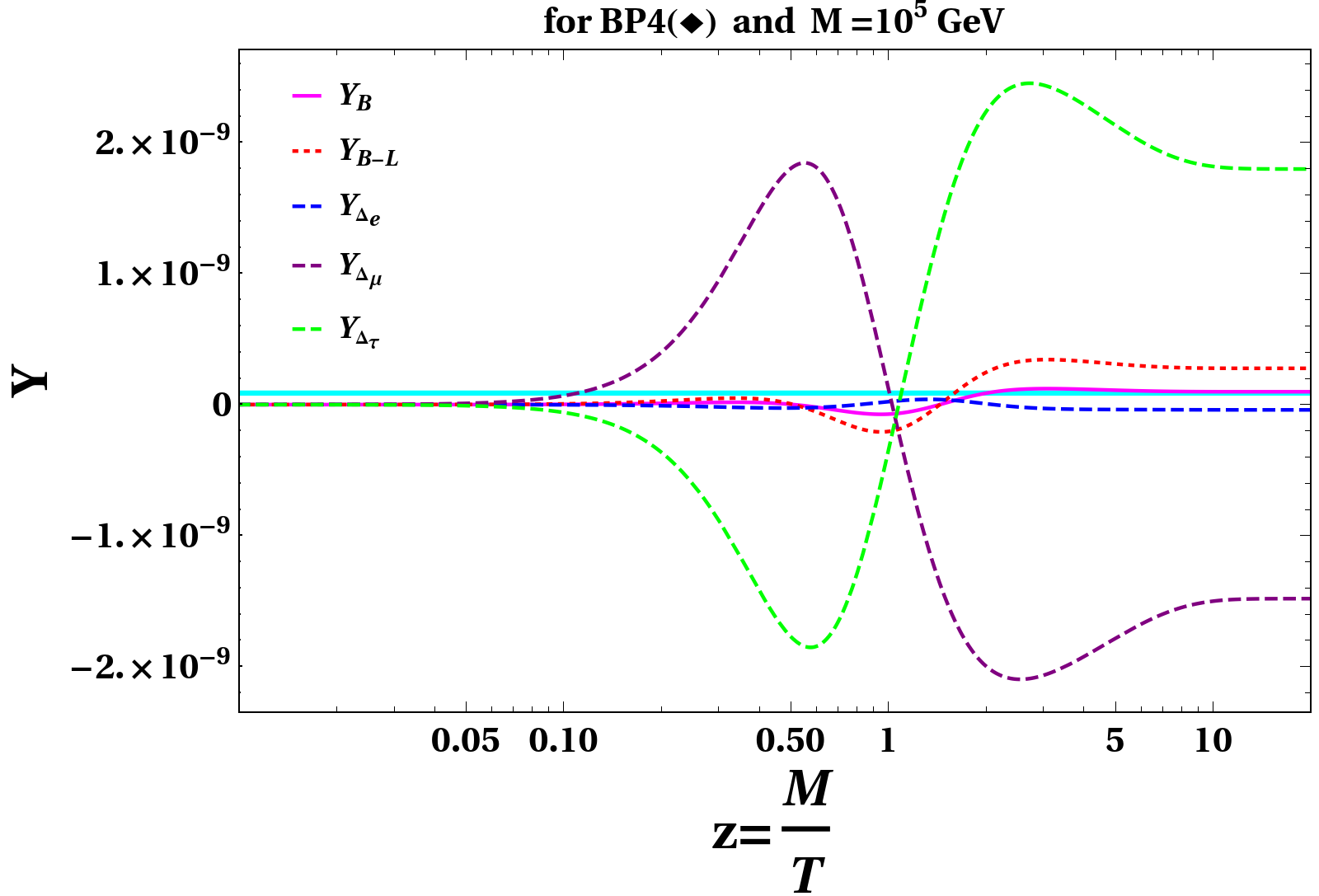}
			\caption{}
			\label{fig:8d}
		\end{subfigure}
		\caption{Variation of $Y_B$, $Y_{B-L}$, $Y_{\Delta_e}$, $Y_{\Delta_{\mu}}$, $Y_{\Delta{\tau}}$ (denoted by solid magenta, dotted red, dashed blue, dashed pink and dashed green lines respectively) presented as function of $z={M}/T$. Here we have considered one benchmark point from each of the four patches of $\gamma_2$ vs $\gamma_1$ plot for the light neutrino parameters of the model (from Fig. \ref{fig:1}).}
		\label{fig:8}
	\end{figure}
	
	With all the ingredients at hand, we first substitute the evaluated CP asymmetry (from Eq. (\ref{epfinal})) in Eq. (\ref{be2}) and proceed for solving the coupled Boltzmann equations in order to find out the final lepton asymmetry as well as final baryon asymmetry. In doing so, we divide the temperature range into three zones so as to take care of the flavor effects 
	as discussed before while taking into account the $\Delta L = 1$ processes (and ignoring $\Delta L = 2$ processes). We have considered different benchmark values for RHN degenerate mass $M$ (splittings are automatically taken cared by running in terms of other parameters): $M= 10^9, 10^6, 10^5$ GeV so that the effects of flavor can be visible. These benchmark values of $M$ are so chosen that they can produce requisite amount of baryon asymmetry corresponding to 
	a specific choice of parameters: $\{ \chi_1, \chi_2, \gamma_1, \gamma_2 \}$. 
	
	\begin{figure}[t]
		\centering
		\includegraphics[width=0.7\linewidth]{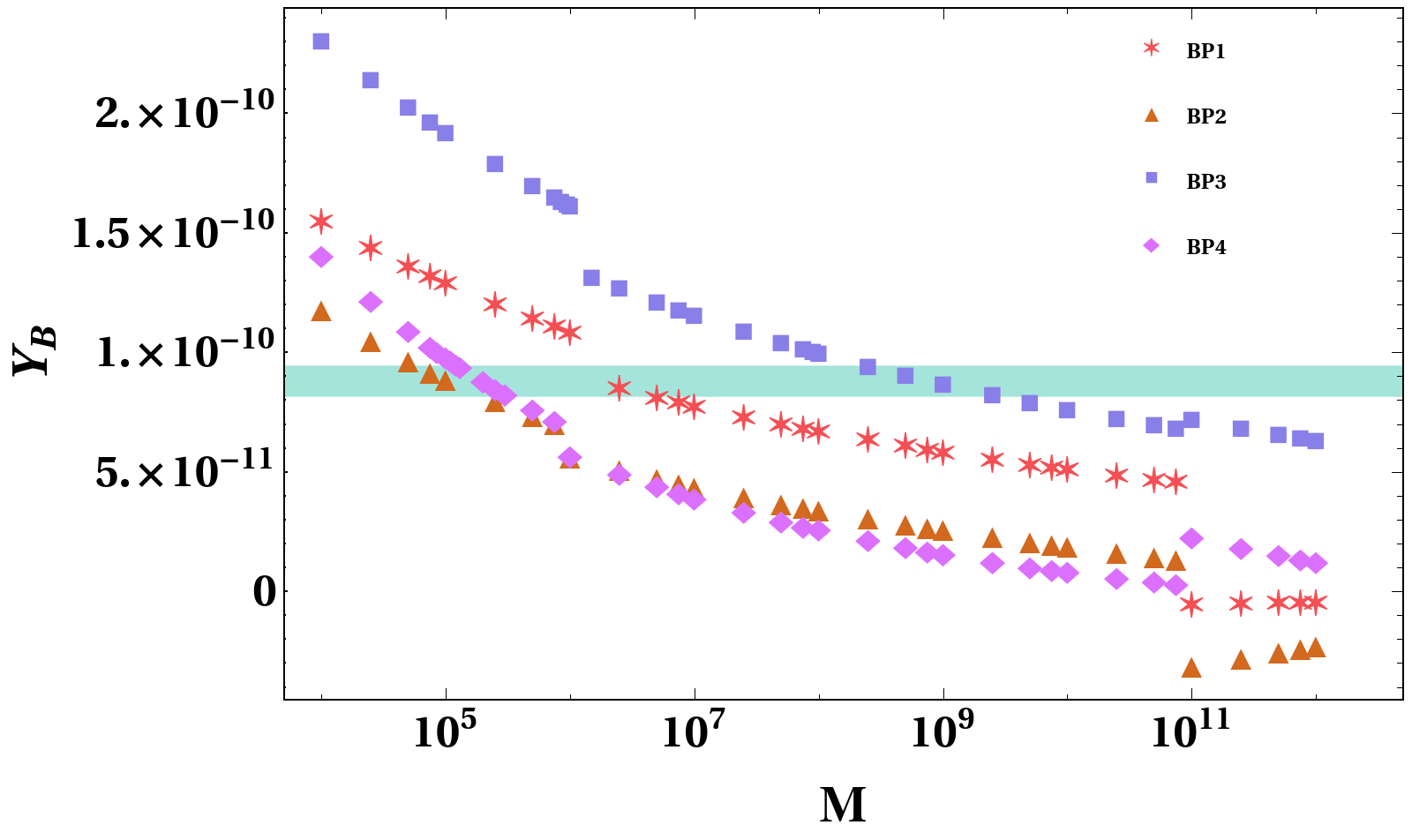}
		\caption{Variation of final $Y_B$ with respect to $M$ (neglecting $\Delta L = 2$ processes) for four  benchmark point BP1, BP2, BP3, BP4, from each of the four patches of $\gamma_2$ vs $\gamma_1$ plot for the light neutrino parameters of the model (from Fig. \ref{fig:1}). Here the horizontal patch (light greenish-blue) indicates the observed value of baryon asymmetry~\cite{Aghanim:2018eyx}.}\label{fig:9}
	\end{figure}
	
	In Fig. \ref{fig:8}, we present our findings in terms of estimate of the evolution of the $B-L$ asymmetry (denoted by red dotted line) as well as $B$ asymmetry (denoted by Magenta solid line) for specific choices of the parameters $\{ \chi_1, \chi_2, \gamma_1, \gamma_2 \}$ which correctly produce neutrino data as discussed in Section \ref{sec:pheno}. Fig. \ref{fig:8a}, \ref{fig:8b},\ref{fig:8c},\ref{fig:8d} represent the benchmark points BP1, BP2, BP3, and BP4 
	respectively from the allowed cornered patches of $\gamma_1$ and $\gamma_2$ plot of Fig. \ref{fig:1}. Asymmetries of individual flavors are also drawn in these figures.

	While solving the Boltzmann equations, we have assumed that initially the abundance of all the RHNs was very less and they were out of equilibrium. Then due to annihilation of bath particles it gets produced and comes to equilibrium. Around $\frac{M}{T} \sim 1$, the production rate and decay rate of the RHN become almost equal and afterward the decay rate dominates over the production rate and hence it's abundance starts to fall. The correct baryon asymmetry 
	can be produced with $M \lesssim 10^6$ GeV for BP1, $M\lesssim 10^5$ GeV for BP2 and BP4, $M \lesssim 10^9 $ GeV for BP3 region respectively. For these individual sets of parameters, we have checked the variation of final baryon asymmetry, $Y_B$, with respect to mass of $M$ as shown in Fig. \ref{fig:9}. From this  Fig. \ref{fig:9}, we also see that final $Y_B$ is increasing with the decreasing of $M$. There seems to be two discontinuities for each such plot. For example, with blue-dotted line, these are observed at or around $M = 10^{11}$ GeV and at $M= 10^6$ GeV. These are indicative of the eras where different flavors of lepton doublets enter in (or exit from) equilibrium and the Boltzmann equations get modified.	
	\section{Conclusion}\label{sec:conc}
	In this analysis, we present an economical, predictive flavor symmetric setup based on $A_4 \times Z_3 \times Z_2$ discrete group to explain neutrino masses, mixing via type-I seesaw mechanism while matter-antimatter asymmetry is also addressed via leptogenesis.  In the original AF model, TBM mixing scheme was realized introducing three flavon fields. With similar fields content, here we show that correct neutrino mixing and mass-squared differences are 
	originated from non-trivial structure of the neutrino Dirac Yukawa coupling and diagonal RHN mass matrix, thanks to the contribution from the charged lepton sector too. In particular, the antisymmetric contribution in the Dirac Yukawa coupling plays an instrumental role in generating the non-zero $\theta_{13}$. Using the current experimental observation on neutrino oscillation and other cosmological limits, we find the allowed parameter space for parameters 
	$\chi_1, \chi_2, \gamma_1, \gamma_2$ which in turn not only restricts some of 
	the observables associated to neutrinos like Dirac CP phase, neutrino-less double beta decay, lepton flavor violating decays, estimation of Majorana phases etc. but also are helpful in determining the matter-antimatter asymmetry of the universe. More specifically, we find that this model is highly predictive in nature. Only normal mass hierarchies are found to be allowed in the current setup. Interestingly the atmospheric mixing angle $\theta_{23}$ lies in the lower octant while the leptonic Dirac CP phase falls within the range $33^{\circ}(213^{\circ})\lesssim \delta \lesssim 80^{\circ}(260^{\circ})$  and  $100^{\circ}(280^{\circ}) \lesssim \delta \lesssim 147^{\circ}(327^{\circ})$. Apart from these predictions for absolute neutrino mass and effective mass parameter appearing the neutrino-less double beta decay have also been made. The model also predicts an interesting correlation between the atmospheric mixing angle $\theta_{23}$ and the Dirac CP phase which is a feature of the specific flavor symmetry considered here. At high scale, owing to the symmetry of the model, the heavy RHNs are found to be exactly degenerate apparently forbidding the generation of baryon asymmetry via leptogenesis. 
	However, this is accomplished here elegantly by considering the renormalization group effects into the picture. A tiny 
	mass splitting produced as a result of running from a high scale (GUT scale) to the scale of RHN mass opens the room for leptogenesis. We have incorporated the flavor effects in leptogenesis as our working regime of RHN mass falls near or below $10^9$ GeV. Finally, we figure out that the parameter space allowed by the neutrino data in fact is good enough to generate sufficient amount of baryon asymmetry of the universe with RHN mass as low as $10^5$ GeV.  
	
	\section*{Acknowledgements}
	The work by BK is supported by the  Polish National Science Centre (NCN) under the Grant  Agreement 2020/37/B/ST2/02371 and DST, Govt. of India (SR/MF/PS-01/2016-IITH/G). BK also acknowledges the support provided by the Institute of High Energy Physics and the University of Chinese Academy of Sciences, Beijing, China, where part of the work has been completed.  AD would like to thank Rishav Roshan and Dibyendu Nanda for fruitful discussions.
	
	\appendix
	\numberwithin{equation}{section}
	\section*{Appendix}
	\section{$A_{4}$ Multiplication Rules:}\label{apa}
	It has four irreducible representations: three one-dimensional and one three dimensional which are denoted by $\bf{1}, 
	\bf{1'}, \bf{1''}$ and $\bf{3}$ respectively. The multiplication rules of the 
	irreducible representations are given by~\cite{Altarelli:2010gt}
	\begin{equation}\label{a4product}
	\bf{1} \otimes \bf{1} = \bf{1}, 
	\bf{1'}\otimes \bf{1'} = \bf{1''},  \bf{1'} \otimes \bf{1''} = \bf{1} ,  \bf{1''} \otimes \bf{1''} = \bf{1'},  \bf{3} \otimes \bf{3} = \bf{1} + \bf{1'} + \bf{1''} + 
	\bf{3}_{a} +    \bf{3}_{s}
	\end{equation}
	where ${\bf a}$ and ${\bf s}$ in the subscript corresponds to anti-symmetric and symmetric 
	parts respectively. Now, if we have two  triplets as $ A = (a_1, a_2, a_3)^T$ and $ B =(b_1, b_2, 
	b_3)^T$ respectively, their direct product can be decomposed into the direct sum 
	mentioned above. The product rule  for this  two triplets  in the $S$ diagonal basis\footnote{Here $S$ is a $3 \times 3$ diagonal generator of $A_4$.} can be written as  
	\begin{eqnarray}
	(A\times B)_{\bf{1}} &\backsim& a_1b_1+a_2 b_2+a_3b_3,\\
	(A\times B)_{\bf{1'}} &\backsim& a_1 b_1 + \omega^2 a_2 b_2 + \omega a_3 b_3,\\
	(A\times B)_{\bf{1''}} &\backsim& a_1 b_1 + \omega a_2 b_2 + \omega^2 a_3 b_3,\\
	(A\times B)_{\bf{3}_{s}} &\backsim& (a_2b_3+a_3b_2,a_3b_1+a_1b_3, a_1b_2+a_2b_1),\label{eq:3s}\\
	(A\times B)_{\bf{3}_{a}} &\backsim& (a_2b_3-a_3b_2, a_3b_1-a_1b_3, a_1b_2-a_2b_1)\label{eq:3a},
	\end{eqnarray}
	here $\omega$ ($=e^{2i\pi/3}$) is the cube root of unity.

	\bibliography{a4reso}
	\bibliographystyle{JHEP}

\end{document}